\documentclass[12pt,twoside, a4paper]{article}
\def\pd{\partial}
\def\mc{\mathcal}

\usepackage[dvips]{graphicx}
\usepackage{amssymb}
\usepackage{amssymb,amsmath}
\usepackage{graphicx}
\usepackage{caption}
\usepackage{subcaption}
\input{epsf.sty}  
\begin{document}
\begin{center}
\LARGE{\textbf{Supersymmetric deformations of 3D SCFTs from
tri-sasakian truncation}}
\end{center}
\vspace{1 cm}
\begin{center}
\large{\textbf{Parinya Karndumri}}
\end{center}
\begin{center}
String Theory and Supergravity Group, Department
of Physics, Faculty of Science, Chulalongkorn University, 254 Phayathai Road, Pathumwan, Bangkok 10330, Thailand
\end{center}
E-mail: parinya.ka@hotmail.com \vspace{1 cm}\\
\begin{abstract}
We holographically study supersymmetric deformations of
$N=3$ and $N=1$ superconformal field theories (SCFTs) in three
dimensions using four-dimensional $N=4$ gauged supergravity coupled
to three-vector multiplets with non-semisimple $SO(3)\ltimes
(\mathbf{T}^3,\hat{\mathbf{T}}^3)$ gauge group. This gauged
supergravity can be obtained from a truncation of eleven-dimensional
supergravity on a tri-sasakian manifold and admits both $N=1,3$
supersymmetric and stable non-supersymmetric $AdS_4$ critical
points. We analyze the BPS equations for $SO(3)$ singlet scalars in
details and study possible supersymmetric solutions. A number of RG
flows to non-conformal field theories and half-supersymmetric domain
walls are found, and many of them can be given analytically. Apart
from these ``flat'' domain walls, we also consider $AdS_3$-sliced
domain wall solutions describing two-dimensional conformal defects
with $N=(1,0)$ supersymmetry within the dual $N=1$ field theory
while this type of solutions does not exist in the $N=3$ case.
\end{abstract}
\newpage
\section{Introduction}
In recent years, superconformal field theories (SCFTs) in three
dimensions have attracted much attention in the context of the
AdS/CFT correspondence \cite{maldacena}. Apart from being effective
world-volume theories of M2-branes \cite{BL,ABJM}, three-dimensional
gauge theories and their conformal fixed points have also
interesting applications in condensed matter physics \cite
{Holographi_Superconductor1,Holographi_Superconductor2,Holographi_Superconductor3}.
\\
\indent Along this line, four-dimensional gauged supergravities have
been a very useful tool in various holographic studies including the
holographic Renormalization Group (RG) flows and conformal defects
of co-dimension one. The former can be described holographically by
domain walls interpolating between two $AdS$ vacua or between an
$AdS$ vacuum in one limit and a domain wall in the other limit, see
for example \cite{fgpw,an,gir}. These two classes of solutions
correspond respectively to RG flows between conformal fixed points
and flows to non-conformal field theories. These solutions are called ``flat'' or Minkowski-sliced domain walls. The conformal defects on
the other hand can be described in the holographic context by
AdS-sliced domain walls
\cite{Bak_Janus,Freedman_Janus,DHoker_Janus,Witten_Janus,Freedman_Holographic_dCFT,5D_Janus_CK,5D_Janus_Suh}.
\\
\indent A number of holographic RG flows within four-dimensional
gauged supergravities have been studied, see for example
\cite{Ahn_4D_flow,Flow_in_N8_4D,4D_G2_flow,Warner_M2_flow,Warner_Fisch} and \cite{Guarino_BPS_DW,Elec_mag_flows,Yi_4D_flow} for  more recent results.
Some of these solutions can be uplifted to eleven dimensions
resulting in many interesting geometric interpretations such as a
polarization of M2-branes into M5-branes in \cite{Warner_N8_uplift}. On the other hand, supersymmetric Janus solutions in four dimensions have been studied
recently in the maximal $N=8$, $SO(8)$ gauged supergravity in
\cite{warner_Janus}. Some of these solutions have been uplifted to
eleven dimensions via a consistent $S^7$ reduction in
\cite{Warner_N8_uplift}. In the context of lower supersymmetry, a
number of supersymmetric Janus solutions within $N=3$, $SU(2)\times
SU(3)$ gauged supergravity have been explored in \cite{N3_Janus}.
This gauged supergravity is expected to describe the lowest Kaluza-Klein modes of a
compactification of M-theory on a tri-sasakian manifold $N^{010}$
\cite{Castellani_Romans}. The gauge group $SU(2)\times SU(3)$ is an
isometry of $N^{010}$, and the two factors are identified with the
$N=3$ $SO(3)_R$ R-symmetry and $SU(3)$ flavor symmetry in the dual SCFT, respectively.
\\
\indent The complete spectrum of this compactification has been
carried out in \cite{N3_spectrum1}, and the structure of the
supermultiplets has been given in \cite{N3_spectrum2}. Furthermore,
the dual SCFT to this compactification has been proposed in
\cite{Ring_N3_superfield}. It has also been discovered in
\cite{Ring_N3_superfield} and further investigated in
\cite{Shadow_N3_multiplet} that all compactifications of M-theory
giving rise to $N=3$ supersymmetric $AdS_4$ backgrounds contain a
universal massive spin-$\frac{3}{2}$ multiplet. All components of
this multiplet arise only from constant harmonics. The truncation
keeping only the lowest Kaluza-Klein modes and this massive multiplet is
accordingly expected to be consistent. The resulting theory is
expected to be $N=4$ gauged supergravity with $N=4$ supersymmetry
broken to $N=3$ at the vacuum. The dual composite operators to
this long, massive, gravitino multiplet have also been proposed in
\cite{Ring_N3_superfield,Shadow_N3_multiplet}.
\\
\indent Up to now, only the complete truncation of
eleven-dimensional supergravity on a generic tri-sasakian manifold
has been carried out in \cite{N010_truncation_Cassani} in which all
the fields which are singlet under the flavor $SU(3)$ symmetry have
been kept. The enhancement by the Betti vector
multiplet, which is also an $SU(3)$ singlet, in the compactification on $N^{010}$ has also
been pointed out. This is due to a non-trivial cohomology of degree two in $N^{010}$ giving rise to an additional massless vector multiplet.
\\
\indent This tri-sasakian truncation results in $N=4$ gauged supergravity coupled to three vector multiplets. The theory admits two supersymmetric
$AdS_4$ solutions with unbroken $SO(3)_R$ R-symmetry and $N=3,1$ supersymmetries. These solutions correspond to
compactifications on $N^{010}$ and its squashed version,
respectively. A possible candidate for the $N=3$ SCFT dual to the $N=3$ solution is given in
\cite{Ring_N3_superfield}, but there is a puzzle with this SCFT regarding to the baryonic spectrum, see a discussion in \cite{Hanany_Zaffaroni1} and \cite{Hanany_Zaffaroni2}. For the $N=1$ case, the situation is less clear. In particular, the $N=1$ SCFT dual to the squashed $N=1$, $AdS_4\times N^{010}$ solution has not previously appeared although the $N=1$ SCFT dual to the squashed $S^7$ compactification has been given in \cite{N1_squashed}. In
this paper, we will analyze the BPS equations for $SO(3)_R$
invariant scalar fields and investigate possible deformations of the
dual $N=3$ and $N=1$ SCFTs within the framework of four-dimensional gauged supergravity.
\\
\indent We will mainly consider supersymmetric
deformations in the forms of RG flows to non-conformal field
theories and two-dimensional defects described by Janus solutions.
Regarding to the $N^{010}$ compactification, a number of holographic RG flows and Janus solutions within the framework
of $N=3$ gauged supergravity have already been studied in
\cite{N3_SU2_SU3} and \cite{N3_Janus}, but these solutions currently
cannot be uplifted to eleven dimensions due to the lack of the complete
consistent truncation keeping all lowest Kaluza-Klein modes including the $SU(3)$ non-singlet ones.
\\
\indent The paper is organized as follow. In section \ref{N4theory},
we review $N=4$ gauged supergravity coupled to three vector
multiplets and the tri-sasakian truncation of eleven-dimensional
supergravity to this $N=4$ gauged supergravity. The
analysis of BPS equations for $SO(3)_R$ singlet scalars will also be carried out. These are relevant for finding supersymmetric RG flow and
Janus solutions in sections \ref{N3_solution} and
\ref{N1_solution}. We will also explicitly give the uplift of some
solutions to eleven dimensions and finally give some conclusions and
comments on the results in section \ref{conclusions}. In the two
appendices, we give an explicit form of the relevant field equations
and some of the complicated BPS equations.

\section{$N=4$ gauged supergravity and tri-sasakian truncation of eleven-dimensional supergravity}\label{N4theory}
In this section, we briefly review $N=4$ gauged supergravity in the
embedding tensor formalism to set up the framework for finding
supersymmetric solutions. Further details on the construction can be
found in \cite{N4_gauged_SUGRA} on which this review is mainly
based. We will also give basic information and relevant formulae of
the tri-sasakian truncation of eleven-dimensional supergravity to
$N=4$ gauged supergravity with $SO(3)\ltimes
(\mathbf{T}^3,\hat{\mathbf{T}}^3)$ gauge group. This is the strategy
we will follow in order to uplift four-dimensional solutions to
eleven dimensions.

\subsection{$N=4$ gauged supergravity coupled to three vector multiplets}
We now consider the half-maximal $N=4$ supergravity in four
dimensions. The supergravity multiplet consists of the graviton
$e^{\hat{\mu}}_\mu$, four gravitini $\psi^i_\mu$, six vectors
$A_\mu^m$, four spin-$\frac{1}{2}$ fields $\chi^i$ and one complex
scalar $\tau$. The complex scalar, or equivalently two real scalars,
can be parametrized by the $SL(2,\mathbb{R})/SO(2)$ coset.
\\
\indent In this half-maximal supersymmetry, the supergravity
multiplet can couple to an arbitrary number $n$ of vector multiplets
although we will later set $n=3$. Each multiplet contains a vector
field $A_\mu$, four gaugini $\lambda^i$ and six scalars $\phi^m$.
The scalar fields can be parametrized by the $SO(6,n)/SO(6)\times
SO(n)$ coset. Before moving to possible gaugings of this
matter-coupled supergravity, we will first give some details on
various indices used throughout this paper.
\\
\indent Space-time and tangent space indices are denoted
respectively by $\mu,\nu,\ldots =0,1,2,3$ and
$\hat{\mu},\hat{\nu},\ldots=0,1,2,3$. The $SO(6)\sim SU(4)$
R-symmetry indices will be described by $m,n=1,\ldots, 6$ for the
$SO(6)$ vector representation and $i,j=1,2,3,4$ for the $SO(6)$
spinor or $SU(4)$ fundamental representations. The $n$ vector
multiplets will be labeled by indices $a,b=1,\ldots, n$. Therefore,
all the fields in the vector multiplets will carry an additional
index in the form of $(A^a_\mu,\lambda^{ia},\phi^{ma})$. All fermionic fields and the supersymmetry parameters transform in the fundamental representation of $SU(4)_R\sim SO(6)_R$ R-symmetry and are subject to the chirality projections
\begin{equation}
\gamma_5\psi^i_\mu=\psi^i_\mu,\qquad \gamma_5\chi^i=-\chi^i,\qquad \gamma_5\lambda^i=\lambda^i\, .
\end{equation}
Similarly, for the corresponding fields transforming in the anti-fundamental representation of $SU(4)_R$, we have
\begin{equation}
\gamma_5\psi_{\mu i}=-\psi_{\mu i},\qquad \gamma_5\chi_i=\chi_i,\qquad \gamma_5\lambda_i=-\lambda_i\, .
\end{equation}
\indent Gaugings of the matter-coupled $N=4$ supergravity can be
efficiently described by using the embedding tensor $\Theta$. This
constant tensor encodes the information about the embedding of any
gauge group $G_0$ in the global or duality symmetry
$SL(2,\mathbb{R})\times SO(6,n)$ in a covariant way. It has been
shown in \cite{N4_gauged_SUGRA} that there are two components of the
embedding tensor $\xi^{\alpha M}$ and $f_{\alpha MNP}$ with
$\alpha=(+,-)$ and $M,N=(m,a)=1,\ldots, n+6$ denoting fundamental
representations of $SL(2,\mathbb{R})$ and $SO(6,n)$, respectively.
The electric vector fields $A^{+M}=(A^m_\mu,A^a_\mu)$, appearing in
the ungauged Lagrangian, and their magnetic dual $A^{-M}$ form a
doublet under $SL(2,\mathbb{R})$ denoted by $A^{\alpha M}$.
\\
\indent In general, a subgroup of both $SL(2,\mathbb{R})$ and
$SO(6,n)$ can be gauged, and the magnetic vector fields can also
participate in the gauging. In particular, it has been shown in \cite{Jan_electricN4},
see also \cite{N4_Wagemans}, that purely electric gaugings do not admit $AdS_4$
vacua. In this paper, we will only consider gaugings involving both
electric and magnetic vector fields in order to obtain $AdS_4$ vacua
relevant for applications in the AdS/CFT correspondence.
\\
\indent The full covariant derivative can be written as
\begin{equation}
D_\mu=\nabla_\mu-gA_\mu^{\alpha M}\Theta_{\alpha M}^{\phantom{\alpha
M}NP}t_{NP}+gA_\mu^{M(\alpha}\epsilon^{\beta)\gamma}\xi_{\gamma
M}t_{\alpha\beta}
\end{equation}
where $\nabla_\mu$ is the usual space-time covariant derivative.
$t_{MN}$ and $t_{\alpha\beta}$ are $SO(6,n)$ and $SL(2,\mathbb{R})$
generators which can be chosen as
\begin{equation}
(t_{MN})_P^{\phantom{P}Q}=2\delta^Q_{[M}\eta_{N]P},\qquad
(t_{\alpha\beta})_\gamma^{\phantom{\gamma}\delta}=2\delta^\delta_{(\alpha}\epsilon_{\beta)\gamma}
\end{equation}
with $\epsilon^{\alpha\beta}=-\epsilon^{\beta\alpha}$ and
$\epsilon^{+-}=1$.
$\eta_{MN}=\textrm{diag}(-1,-1,-1,-1,-1,-1,1,\ldots,1)$ is the
$SO(6,n)$ invariant tensor, and $g$ is the gauge coupling constant
that can be absorbed in the embedding tensor $\Theta$. The embedding
tensor appearing in the above equation can be written in terms of
$\xi^{\alpha M}$ and $f_{\alpha MNP}$ as
\begin{equation}
\theta_{\alpha MNP}=f_{\alpha MNP}-\xi_{\alpha[N}\eta_{P]M}\, .
\end{equation}
In the following discussions, we will only consider solutions with
only the metric and scalars non-vanishing. Therefore, we will set
all of the vector fields to zero from now on.
\\
\indent We now consider explicit parametrization of the scalar coset
manifold $SL(2,\mathbb{R})/SO(2)\times SO(6,n)/SO(6)\times SO(n)$.
The first factor can be described by a coset representative
\begin{equation}
\mc{V}_\alpha=\frac{1}{\sqrt{\textrm{Im} \tau}}\left(
                                         \begin{array}{c}
                                           \tau \\
                                           1 \\
                                         \end{array}
                                       \right)
\end{equation}
or equivalently by a symmetric matrix
\begin{equation}
M_{\alpha\beta}=\textrm{Re} (\mc{V}_\alpha\mc{V}^*_\beta)=\frac{1}{\textrm{Im}
\tau}\left(
                                    \begin{array}{cc}
                                      |\tau|^2 & \textrm{Re} \tau \\
                                      \textrm{Re} \tau & 1 \\
                                    \end{array}
                                  \right).
\end{equation}
Note that $\textrm{Im}(\mc{V}_\alpha\mc{V}^*_\beta)=\epsilon_{\alpha\beta}$.
The complex scalar $\tau$ can in turn be written in terms of the
dilaton $\phi$ and the axion $\chi$ as
\begin{equation}
\tau=\chi+ie^\phi\, .
\end{equation}
\indent For the $SO(6,n)/SO(6)\times SO(n)$ factor, we introduce the
coset representative $\mc{V}_M^{\phantom{M}A}$ transforming by a
left and right multiplication under $SO(6,n)$ and $SO(6)\times
SO(n)$, respectively. We will split the $SO(6)\times SO(n)$ index
$A=(m,a)$ and write the coset representative as
$\mc{V}_M^{\phantom{M}A}=(\mc{V}_M^{\phantom{M}m},\mc{V}_M^{\phantom{M}a})$.
Being an element of $SO(6,n)$, the matrix $\mc{V}_M^{\phantom{M}A}$
satisfies the relation
\begin{equation}
\eta_{MN}=-\mc{V}_M^{\phantom{M}m}\mc{V}_N^{\phantom{M}m}+\mc{V}_M^{\phantom{M}a}\mc{V}_N^{\phantom{M}a}\,
.
\end{equation}
As in the $SL(2,\mathbb{R})/SO(2)$ factor, we can parametrize the
$SO(6,n)/SO(6)\times SO(n)$ coset in term of a symmetric matrix
\begin{equation}
M_{MN}=\mc{V}_M^{\phantom{M}m}\mc{V}_N^{\phantom{M}m}+\mc{V}_M^{\phantom{M}a}\mc{V}_N^{\phantom{M}a}\,
.
\end{equation}
\indent We are now in a position to give the bosonic Lagrangian with
the vector fields and auxilary two-form fields vanising
\begin{equation}
e^{-1}\mc{L}=\frac{1}{2}R+\frac{1}{16}\pd_\mu M_{MN}\pd^\mu
M^{MN}-\frac{1}{4(\textrm{Im}\tau)^2}\pd_\mu \tau \pd^\mu \tau^*-V
\end{equation}
where $e$ is the vielbein determinant. The scalar potential is given
by
\begin{eqnarray}
V&=&\frac{g^2}{16}\left[f_{\alpha MNP}f_{\beta
QRS}M^{\alpha\beta}\left[\frac{1}{3}M^{MQ}M^{NR}M^{PS}+\left(\frac{2}{3}\eta^{MQ}
-M^{MQ}\right)\eta^{NR}\eta^{PS}\right]\right.\nonumber \\
& &\left.-\frac{4}{9}f_{\alpha MNP}f_{\beta
QRS}\epsilon^{\alpha\beta}M^{MNPQRS}+3\xi^M_\alpha \xi^N_\beta
M^{\alpha\beta}M_{MN}\right]
\end{eqnarray}
where $M^{MN}$ is the inverse of $M_{MN}$, and $M^{MNPQRS}$ is
defined by
\begin{equation}
M_{MNPQRS}=\epsilon_{mnpqrs}\mc{V}_{M}^{\phantom{M}m}\mc{V}_{N}^{\phantom{M}n}
\mc{V}_{P}^{\phantom{M}p}\mc{V}_{Q}^{\phantom{M}q}\mc{V}_{R}^{\phantom{M}r}\mc{V}_{S}^{\phantom{M}s}\label{M_6}
\end{equation}
with indices raised by $\eta^{MN}$.
\\
\indent The gauge group we will consider here is a non-semisimple
group $SO(3)\ltimes (\mathbf{T}^3,\hat{\mathbf{T}}^3)\subset
SO(6,3)$ described by the non-vanishing component $f_{\alpha MNP}$
of the embedding tensor. We will then set $\xi^{\alpha M}=0$ in the
following discussion. The embedding of this $SO(3)\ltimes
(\mathbf{T}^3,\hat{\mathbf{T}}^3)$ gauge group is described by the
following components of the embedding tensor
\begin{eqnarray}
f_{+IJ,K+6}&=&-f_{+I+3,J+6,K+6}=-2\sqrt{2}\epsilon_{IJK},\qquad I,J,K=1,2,3,\nonumber \\
f_{+I+6,J+6,K+6}&=&6\sqrt{2}k\epsilon_{IJK},\qquad f_{-I,J+6,K+6}=-4\epsilon_{IJK}\, .\label{embedding_tensor}
\end{eqnarray}
The constant $k$ is related to the four-form flux along the
four-dimensional space-time, see equation \eqref{4_form_flux} below.
This gauge group arises from a truncation of eleven-dimensional
supergravity on a tri-sasakain manifold
\cite{N010_truncation_Cassani}. It should be noted that both
electric and magnetic components participate in the gauging, $f_{\pm
MNP}\neq 0$, since purely electric gaugings do not lead to $AdS_4$
vacua as mentioned above.
\\
\indent We should also remark that the identification of this gauge
group and other computations in \cite{N010_truncation_Cassani} have
been done in the off-diagonal $\eta_{MN}$
\begin{equation}
\eta_{MN}=\left(
                     \begin{array}{ccc}
                       -\mathbf{I}_3 & \mathbf{0}_3 & \mathbf{0}_3 \\
                       \mathbf{0}_3 & \mathbf{0}_3 & \mathbf{I}_3 \\
                       \mathbf{0}_3 & \mathbf{I}_3 & \mathbf{0}_3 \\
                     \end{array}
                   \right)
\end{equation}
where $\mathbf{0}_3$ and $\mathbf{I}_3$ denote $3\times 3$ zero and
identity matrices, respectively. Accordingly, in computing
$M_{MNPQRS}$ in \eqref{M_6} and some parts of the supersymmetry
transformations given below, $\mc{V}_M^{\phantom{M}m}$ and
$\mc{V}_M^{\phantom{M}a}$ must be projected to the negative and
positive eigenvalue subspaces of $\eta_{MN}$, respectively.
\\
\indent By transforming to a purely electric frame, the gauge
algebra will be more transparent. We will not explicitly give this
transformation here since we will mainly work in the above
electric-magnetic frame. However, for completeness, we will discuss
the structure of the gauge algebra here, see
\cite{N010_truncation_Cassani} for more details. The $SO(3)$ part is
the diagonal subgroup of $SO(3)\times SO(3)\times SO(3)\subset
SO(6)\times SO(3)\subset SO(6,3)$. The six generators of
$\mathbf{T}^3$ and $\hat{\mathbf{T}}^3$ transform as
$\mathbf{3}+\mathbf{3}$ under $SO(3)$. $\mathbf{T}^3$ generators
commute with each other while $\hat{\mathbf{T}}^3$ generators close
on to $\mathbf{T}^3$ generators.
\\
\indent We now turn to another important ingredient of the $N=4$
gauged supergravity namely the supersymmetry transformations of
fermionic fields. These are given by
\begin{eqnarray}
\delta\psi^i_\mu &=&2D_\mu \epsilon^i-\frac{2}{3}gA^{ij}_1\gamma_\mu
\epsilon_j,\\
\delta \chi^i &=&i\epsilon^{\alpha\beta}\mc{V}_\alpha D_\mu
\mc{V}_\beta\gamma^\mu \epsilon^i-\frac{4}{3}igA_2^{ij}\epsilon_j,\\
\delta \lambda^i_a&=&2i\mc{V}_a^{\phantom{a}M}D_\mu
\mc{V}_M^{\phantom{M}ij}\gamma^\mu\epsilon_j+2igA_{2aj}^{\phantom{2aj}i}\epsilon^j\,
.
\end{eqnarray}
The fermion shift matrices are defined by
\begin{eqnarray}
A_1^{ij}&=&\epsilon^{\alpha\beta}(\mc{V}_\alpha)^*\mc{V}_{kl}^{\phantom{kl}M}\mc{V}_N^{\phantom{N}ik}
\mc{V}_P^{\phantom{P}jl}f_{\beta M}^{\phantom{\beta M}NP},\nonumber
\\
A_2^{ij}&=&\epsilon^{\alpha\beta}\mc{V}_\alpha\mc{V}_{kl}^{\phantom{kl}M}\mc{V}_N^{\phantom{N}ik}
\mc{V}_P^{\phantom{P}jl}f_{\beta M}^{\phantom{\beta M}NP},\nonumber
\\
A_{2ai}^{\phantom{2ai}j}&=&\epsilon^{\alpha\beta}\mc{V}_\alpha
\mc{V}^M_{\phantom{M}a}\mc{V}^N_{\phantom{N}ik}\mc{V}_P^{\phantom{P}jk}f_{\beta
MN}^{\phantom{\beta MN}P}
\end{eqnarray}
where $\mc{V}_M^{\phantom{M}ij}$ is defined in terms of the 't Hooft
symbols $G^{ij}_m$ and $\mc{V}_M^{\phantom{M}m}$ as
\begin{equation}
\mc{V}_M^{\phantom{M}ij}=\frac{1}{2}\mc{V}_M^{\phantom{M}m}G^{ij}_m
\end{equation}
an similarly for its inverse
\begin{equation}
\mc{V}^M_{\phantom{M}ij}=-\frac{1}{2}\mc{V}_M^{\phantom{M}m}(G^{ij}_m)^*\,
.
\end{equation}
$G^{ij}_m$ satisfy the relations
\begin{equation}
G_{mij}=(G^{ij}_m)^*=\frac{1}{2}\epsilon_{ijkl}G^{kl}_m\, .
\end{equation}
The explicit form of these matrices can be found for example in
\cite{Jan_electricN4}. Note that we use the convention about the
(anti) self-duality of $G_{mij}$ opposite to that of
\cite{Jan_electricN4}. It should also be noted that the scalar
potential can be written in terms of $A_1$ and $A_2$ tensors as
\begin{equation}
V=-\frac{1}{3}A^{ij}_1A_{1ij}+\frac{1}{9}A^{ij}_2A_{2ij}+\frac{1}{2}A_{2ai}^{\phantom{2ai}j}
A_{2a\phantom{i}j}^{\phantom{2a}i}\, .
\end{equation}

\subsection{$N=4$ gauged supergravity from eleven dimensions}
Four-dimensional $N=4$ gauged supergravity coupled to three vector
multiplets with $SO(3)\ltimes (\mathbf{T}^3,\hat{\mathbf{T}}^3)$
gauge group has been obtained from a truncation of
eleven-dimensional supergravity on a generic tri-sasakian manifold
in \cite{N010_truncation_Cassani}. In this section, we review
relevant formulae involving the reduction ansatz which will be
useful for uplifting four-dimensional solutions in the next
sections. In particular, we will set all of the vector fields to
zero as well as the auxiliary two-form and magnetic vector fields.
\\
\indent The eleven-dimensional metric can be written as
\begin{equation}
ds^2_{11}=e^{2\varphi}ds^2_4+e^{2U}ds^2(B_\textrm{QK})+g_{IJ}\eta^I\eta^J\,
.
\end{equation}
The three-dimensional internal metric $g_{IJ}$ can be written in
terms of the vielbein as
\begin{equation}
g=Q^TQ\, .
\end{equation}
For convenience, as in \cite{N010_truncation_Cassani}, we will parametrize the matrix $Q$ in
term of a product of a diagonal matrix $V$ and an $SO(3)$ matrix $O$
as
\begin{equation}
Q=VO,\qquad V=\textrm{diag}(e^{V_1},e^{V_2},e^{V_3})\, .
\end{equation}
The scalar $\varphi$ is chosen in such a way that the
four-dimensional Einstein-Hilbert term is obtained
\begin{equation}
\varphi=-\frac{1}{2}(4U+V_1+V_2+V_3).
\end{equation}
Finally, $B_{\textrm{QK}}$ denotes a four-dimensional quaternionic
Kahler manifold.
\\
\indent The three-form field and its four-form field strength are given respectively by
\begin{equation}
C_3=c_3+c_{IJ}\eta^I\wedge J^I+\frac{1}{6}\chi \epsilon_{IJK}\eta^I\wedge \eta^J\wedge \eta^K
\end{equation}
and
\begin{eqnarray}
G_4&=&H_4+4\textrm{Tr}c\,\textrm{vol}(\textrm{QK})+\frac{1}{6}\epsilon_{IJK}d\chi\wedge\eta^I\eta^J\eta^K
+dc_{IJ}\wedge\eta^I \wedge J^J\nonumber \\
& &\epsilon_{IJL}\left[(\chi+\textrm{Tr}c)\delta_{LK}-2c_{(LK)}\right]\eta^I\wedge\eta^J\wedge J^K
\end{eqnarray}
where $H_4=dc_3$, $c_{IJ}$ is a $3\times 3$ matrix and $\textrm{Tr}
c=\delta^{IJ}c_{IJ}$. In the present case, the $H_4$ will be given
by
\begin{equation}
H_4=-6ke^{4\varphi-V_1-V_2-V_3-4U}\textrm{vol}_4\label{4_form_flux}
\end{equation}
where $\textrm{vol}_4$ is the volume form of the four-dimensional
metric $ds^2_4$. The volume form of $B_{\textrm{QK}}$,
$\textrm{vol}(\textrm{QK})$, can be written in terms of the two-forms $J^I$
as
\begin{equation}
\textrm{vol}(\textrm{QK})=\frac{1}{6}J^I\wedge J^I\, .
\end{equation}
\indent For the $N^{010}$ tri-sasakian manifold, we can take a
simple description in term of a coset manifold $SU(3)/U(1)$. This is
enough for our propose although the full $SU(3)\times SU(2)$
isometry is not manifest, see \cite{N010_Castellani} for another
description. Using the standard Gell-Mann matrices, we can choose
the $SU(3)$ geneartors to be $-\frac{i}{2}\lambda_\alpha$,
$\alpha=1,\ldots, 8$. The coset and $U(1)$ generators can be chosen
to be
\begin{equation}
K_i=-\frac{i}{2}(\lambda_1,\lambda_2,\lambda_3,\lambda_4,\lambda_5,\lambda_6,\lambda_7),\qquad
H=-\frac{i\sqrt{3}}{2}\lambda_8\, .
\end{equation}
The vielbein on $N^{010}$ can eventually be obtained from the decomposition of the Maurer-Cartan one-form
\begin{equation}
L^{-1}dL=e^iK_i+\omega H
\end{equation}
where $L$ is the coset representative for $SU(3)/U(1)$. $\omega$ is
the corresponding $U(1)$ connection.
\\
\indent Following \cite{N010_truncation_Cassani}, we will use the tri-sasakian structures of the form
\begin{eqnarray}
\eta^I&=&\frac{1}{2}(e^1,e^2,e^3),\nonumber\\
J^I&=&\frac{1}{8}(e^4\wedge e^5-e^3\wedge e^6,-e^3\wedge e^5-e^4\wedge e^6,e^5\wedge e^6-e^3\wedge e^4).
\end{eqnarray}
From these, we find the metric on $B_{\textrm{QK}}$ to be
\begin{equation}
ds^2(B_{\textrm{QK}})=\frac{1}{256}\left[(e^3)^2+(e^4)^2+(e^5)^2+(e^6)^2\right]
\end{equation}
with the volume form given by
\begin{equation}
\textrm{vol}(\textrm{QK})=\frac{1}{6}J^I\wedge J^I=-\frac{1}{64}e^3\wedge e^4\wedge e^5\wedge e^6\, .
\end{equation}
In the remaining parts of this paper, we will not need the explicit
form of $ds^2(B_{\textrm{QK}})$ and $\eta^I$'s since we will not
consider the deformations of these metrics. Therefore, we will leave these as generic expressions.

\subsection{BPS equations for $SO(3)$ invariant scalars}
We now give an explicit parametrization of the
$SL(2,\mathbb{R})/SO(2)\times SO(6,3)/SO(6)\times SO(3)$ coset and
relevant information for setting up the BPS equations corresponding
to $SO(3)$ singlet scalars.
\\
\indent Since we will study both RG flows and Janus solutions, and
the former can formally be obtained as a limit of the latter, we
will first construct the BPS equations for finding supersymmetric
Janus solutions and take an appropriate limit to find the BPS
equations for RG flow solutions. The metric ansatz takes the form of
an $AdS_3$-sliced domain wall
\begin{equation}
ds^2=e^{2A(r)}\left(e^{\frac{2\xi}{\ell}}dx^2_{1,1}+d\xi^2\right)+dr^2\,
.\label{DW_ansatz}
\end{equation}
As can be clearly seen, this metric becomes a flat domain wall used
in the study of holographic RG flows in the limit $\ell\rightarrow
\infty$. The vielbein components can be chosen to be
\begin{equation}
e^{\hat{\mu}}=e^{A+\frac{\xi}{\ell}}dx^\mu,\qquad
e^{\hat{\xi}}=e^{A}d\xi,\qquad e^{\hat{r}}=dr\, .
\end{equation}
The non-vanishing spin connections of this metric are then given by
\begin{equation}
\omega^{\hat{\xi}}_{\phantom{\hat{\xi}}\hat{r}}=A'e^{\hat{\xi}},\qquad \omega^{\hat{\mu}}_{\phantom{\hat{\xi}}\hat{\xi}}=\frac{1}{\ell}e^{-A}e^{\hat{\mu}},\qquad
\omega^{\hat{\mu}}_{\phantom{\hat{\xi}}\hat{r}}=A'e^{\hat{\mu}}
\end{equation}
where $'$ denotes the $r$-derivative. For the moment, indices $\mu,\nu$
will take values $0,1$, and hatted indices are the tangent space
indices.
\\
\indent In this paper, we are only interested in $SO(3)$ singlet scalars. These
scalar fields depend only on the radial coordinate $r$. There are
four $SO(3)$ singlets corresponding to two scalars from
$SL(2,\mathbb{R})/SO(2)$ and another two from $SO(6,3)/SO(6)\times
SO(3)$ according to the branching of $SO(6,3)\rightarrow SO(3)\times
SO(3)\times SO(3)\rightarrow SO(3)_{\textrm{diag}}$
\begin{equation}
(\mathbf{6},\mathbf{3})\rightarrow
(\mathbf{3},\mathbf{1},\mathbf{3})+(\mathbf{1},\mathbf{3},\mathbf{3})\rightarrow
2\times (\mathbf{1}+\mathbf{3}+\mathbf{5}).
\end{equation}
Following \cite{N010_truncation_Cassani}, we parametrize the
$SO(6,3)/SO(6)\times SO(3)$ coset representative by
\begin{equation}
\mc{V}=\textrm{exp}\left(
                     \begin{array}{ccc}
                       \mathbf{0}_3 & \sqrt{2}Z\mathbf{I}_3 & \mathbf{0}_3 \\
                       \mathbf{0}_3 & \mathbf{0}_3 & \mathbf{0}_3 \\
                       \sqrt{2}Z\mathbf{I}_3 & \mathbf{0}_3 & \mathbf{0}_3 \\
                     \end{array}
                   \right)\times \left(
                                   \begin{array}{ccc}
                                     \mathbf{I}_3 & \mathbf{0}_3 & \mathbf{0}_3 \\
                                     \mathbf{0}_3 & e^{-2U-V_1}\mathbf{I}_3 & \mathbf{0}_3 \\
                                     \mathbf{0}_3 & \mathbf{0}_3 & e^{2U+V_1}\mathbf{I}_3 \\
                                   \end{array}
                                 \right).\label{SO3_inv_coset}
\end{equation}
Note that $SO(3)$ invariance requires $c_{IJ}$ to be proportional to
the identity, $c_{IJ}=\sqrt{2}Z\delta_{IJ}$, and $V_1=V_2=V_3$.
\\
\indent The $SL(2,\mathbb{R})/SO(2)$ scalars are given by
\begin{equation}
\tau=\chi+ie^{3V_1}\, .
\end{equation}
For convenience, we will define another scalar
\begin{equation}
U_1=2U+V_1\, .
\end{equation}
This also gives a diagonal scalar kinetic term
\begin{equation}
\frac{1}{16}\pd_\mu M_{MN}\pd^\mu M^{MN}-\frac{1}{4 (\textrm{Im}
\tau)^2}\pd_\mu \tau \pd^\mu \tau^*=
-\frac{3}{2}{U_1'}^2-\frac{9}{4}{V_1'}^2-\frac{1}{4}e^{-6V_1}{\chi'}^2-\frac{3}{2}e^{-2U_1}{Z'}^2\,
.\label{scalar_kin}
\end{equation}
\indent In order to setup the BPS equations corresponding to $\delta
\chi^i=0$ and $\delta\lambda^i_a=0$, a projector involving
$\gamma_r$ is needed. Since the procedure is essentially the same as
in \cite{warner_Janus} and \cite{N3_Janus}, we will only repeat the
relevant formulae. Following \cite{warner_Janus}, we will use
Majorana representation in which all gamma matrices $\gamma_\mu$ are
real, and
$\gamma_5=i\gamma_{\hat{0}}\gamma_{\hat{1}}\gamma_{\hat{\xi}}\gamma_{\hat{r}}$
is purely imaginary. In the chiral notation, we have, for example,
\begin{equation}
\epsilon^i=\frac{1}{2}(1+\gamma_5)\epsilon^i_M,\qquad \epsilon_i=\frac{1}{2}(1-\gamma_5)\epsilon^i_M
\end{equation}
where $\epsilon_M$ is a four-component Majorana spinor. From these, it follows that $\epsilon_i=(\epsilon^i)^*$.
\\
\indent Accordingly, the $\gamma_r$-projector can be written as
\begin{equation}
\gamma^{\hat{r}}\epsilon^i=e^{i\Lambda}\epsilon_i
\end{equation}
or equivalently
\begin{equation}
\gamma^{\hat{r}}\epsilon_i=e^{-i\Lambda}\epsilon^i\, .
\end{equation}
The analysis of $\delta \psi^i_\mu=0$ equations leads to the following $\gamma_{\hat{\xi}}$ projection
\begin{equation}
\gamma_{\hat{\xi}}\epsilon_i=i\kappa e^{i\Lambda}\epsilon^i
\end{equation}
see \cite{warner_Janus} for more detail. The constant $\kappa$ satisfying $\kappa^2=1$ determines the chirality of the unbroken supercharges on the two-dimensional defect. Up to a phase, the full Killing spinor can be written as
\begin{equation}
\epsilon^i=e^{\frac{A}{2}+\frac{\xi}{2\ell}+i\frac{\Lambda}{2}}\varepsilon^{(0)i}
\end{equation}
with the constant spinors $\varepsilon^{(0)i}$ satisfying
\begin{equation}
\gamma_{\hat{r}}\varepsilon^{(0)i}=\varepsilon^{(0)}_i\qquad
\textrm{and}\qquad
\gamma_{\hat{\xi}}\varepsilon^{(0)}_i=i\kappa\varepsilon^{(0)i}\, .
\end{equation}
\indent The integrabitity conditions of $\delta
\psi^i_{\hat{0},\hat{1}}=0$ equations give
\begin{equation}
A'^2+\frac{1}{\ell^2}e^{-2A}=|\mc{W}|^2
\end{equation}
where $\mc{W}$ is the ``superpotential'' given by the eigenvalue
$\alpha$ of the $A_1^{ij}$ tensor corresponding to the unbroken
supersymmetry
\begin{equation}
\mc{W}=\frac{2}{3}\alpha\, .
\end{equation}
The cosmological constant at $AdS_4$ critical points is given in term of $\alpha$ by the relation $V_0=-\frac{4}{3}\alpha^2$.
\\
\indent Finally, we note the expression for the phase $e^{i\Lambda}$ in terms of $\mc{W}$
\begin{eqnarray}
e^{i\Lambda}&=&\frac{A'}{W}+\frac{i\kappa}{\ell}\frac{e^{-A}}{W}\, .\label{real_W_phase}\\
\textrm{and}\qquad
e^{i\Lambda}&=&\frac{\mc{W}}{A'+\frac{i\kappa}{\ell}e^{-A}}\, .\label{complex_W_phase}
\end{eqnarray}
for real and complex $\mc{W}$, respectively. These relations can be
obtained by considering the gravitino variations in each case, see
\cite{N3_Janus} for more detail.
\\
\indent For the RG flows, the corresponding BPS equations can be found by formally taking the limit $\ell\rightarrow \infty$. We simply find
\begin{equation}
A'=\pm W\qquad \textrm{and}\qquad e^{i\Lambda}=\frac{\mc{W}}{W}\, .
\end{equation}
where $W=|\mc{W}|$ is call the ``real superpotential''. The
$\gamma_{\hat{\xi}}$ projector drops out, and there is no chirality
restriction on the preserved supercharges.
\\
\indent We now give the scalar potential for $SO(3)$ singlet scalars
\begin{eqnarray}
V&=&3e^{-6U_1-3V_1}\left[2e^{2U_1+6V_1}+12e^{6V_1}Z^2-e^{4U_1}-8e^{3(U_1+V_1)} \right. \nonumber \\
& &\left. +2e^{2U_1}(\chi+Z)^2+3(k-2\chi
Z-Z^2)^2\right].\label{Scalar_potential}
\end{eqnarray}
As pointed out in \cite{N010_truncation_Cassani}, this is the scalar
potential of the truncated $N=1$ supergravity in which only $SO(3)$
singlet fields are retained.
\\
\indent The scalar field equations can be obtained by using this
potential in the effective Lagrangian
\begin{equation}
\mc{L}_{\textrm{scalar}}=e^{3A}\left[\frac{1}{16}\pd_\mu M_{MN}\pd^\mu
M^{MN}-\frac{1}{4(\textrm{Im}\tau)^2}\pd_\mu \tau \pd^\mu
\tau^*-V\right].
\end{equation}
Note that the scalar field equations are the same for both the RG
flows and Janus solutions since scalars do not depend on the $\xi$
coordinate. This is the reason we can take $\sqrt{-g}$ to be just
$e^{3A}$ not $e^{3A+2\frac{\xi}{\ell}}$. The explicit form of these
equations and Einstein equations will be given in appendix
\ref{field_eq}.
\\
\indent As shown in \cite{N010_truncation_Cassani}, the above
potential admits a number of $AdS_4$ critical points both
supersymmetric and non-supersymmetric. In this paper, we will only
consider the following supersymmetric $AdS_4$ vacua
\begin{eqnarray}
\textrm{I}&:&\quad U_1=3V_1=\frac{1}{2}\ln |k|,\qquad V_0=-12|k|^{-\frac{3}{2}}\\
\textrm{II}&:&\quad U_1=\ln 5+\frac{1}{2}\ln \frac{|k|}{15},\qquad
V_1=\frac{1}{6}\ln \frac{|k|}{15},\qquad
V_0=-12|k|^{-\frac{3}{2}}\sqrt{\frac{3^7}{5^5}}
\end{eqnarray}
with $\chi=Z=0$. The cosmological constant $V_0$ is related to the $AdS_4$ radius by
\begin{equation}
L^2=-\frac{3}{V_0}\, .
\end{equation}
Within the $N=4$ gauged supergravity, critical point I with $k>0$
gives $N=3$ supersymmetric $AdS_4$ vacuum while $k<0$ solution gives
a non-supersymmetric skew-whiffle solution as will be shown in the
next section. Similarly, critical point II with $k<0$ and $k>0$
corresponds respectively to weak $G_2$ $N=1$ $AdS_4$ and
non-supersymmetric skew-whiffle solutions. In particular, the $N=1$
critical point corresponds to a squashed version of $N^{010}$
manifold. It is also useful to note the two metrics here
\begin{eqnarray}
N=3:\quad
ds^2_{11}&=&|k|^{-\frac{7}{6}}\left(e^{\frac{2r}{L_3}}dx^2_{1,2}+dr^2\right)+|k|^{\frac{1}{3}}
\left[ds^2(B_{\textrm{QK}})+\eta^I\eta^I\right], \\
N=1:\quad
ds^2_{11}&=&\frac{1}{25}\left(\frac{|k|}{15}\right)^{-\frac{7}{6}}\left(e^{\frac{2r}{L_1}}dx^2_{1,2}+dr^2\right)+5\left(\frac{|k|}{15}\right)^{\frac{1}{3}}
\left[ds^2(B_{\textrm{QK}})+\frac{1}{5}\eta^I\eta^I\right]\nonumber \\
& &
\end{eqnarray}
where the $AdS_4$ radii are given by
$L_3=\frac{1}{2}|k|^{\frac{3}{4}}$ and
$L_1=\frac{5^{\frac{5}{4}}}{2(3)^{\frac{7}{4}}}|k|^{\frac{3}{4}}$.
\\
\indent Before carrying out the analysis of BPS equations, we
briefly discuss the dual SCFTs to these critical points. The SCFT
dual to the $N=3$ critical point has been proposed in
\cite{Ring_N3_superfield}. At low energy, this is an $SU(N)\times
SU(N)$ gauge theory of interacting three hypermultiplets
transforming in a triplet of the $SU(3)$ flavor symmetry. Each
hypermultiplet transforms as a bifundamental under the $SU(N)\times
SU(N)$ gauge group and as a doublet of the $SU(2)_R\sim SO(3)_R$
R-symmetry. In terms of the $N=2$ superfields, these hypermultiplets
can be written as
\begin{equation}
U^i_\alpha=(u^i,-\bar{v}^i)\qquad \textrm{and}\qquad
V_{i\alpha}=-\epsilon_{\alpha\beta}\bar{U}^\beta_i=(v_i,\bar{u}_i)
\end{equation}
where $i=1,2,3$ and $\alpha=1,2$.
\\
\indent From the Kaluza-Klein spectrum given in \cite{N3_spectrum1}
and \cite{N3_spectrum2}, the massless graviton multiplet corresponds
to the usual stress-energy tensor multiplet, including the $SO(3)_R$
R-symmetry current, in the dual $N=3$ SCFT. There are also nine
massless vector multiplets transforming in the adjoint and singlet
(Betti multiplet) representations of $SU(3)$. These correspond to
the following operator
\begin{eqnarray}
\Sigma^i_{\phantom{i}j}&=&\frac{1}{\sqrt{2}}\textrm{Tr}(U^i\bar{U}_j+\bar{V}^iV_j)
-\frac{1}{3\sqrt{2}}\delta^i_j\textrm{Tr}(U^k\bar{U}_k+\bar{V}^kV_k),\\
\Sigma&=&\frac{1}{\sqrt{2}}\textrm{Tr}(U^i\bar{U}_i+\bar{V}^iV_i)
\end{eqnarray}
which are the conserved currents of the flavor $SU(3)$ and the
baryonic $U(1)$ global symmetries, respectively.
\\
\indent In \cite{Ring_N3_superfield}, see also
\cite{Shadow_N3_multiplet}, the operator dual to the massive
gravitino multiplet, which is of particular interest in the present
work, has also been proposed. The corresponding operator is given by
the $SO(3)_R$ singlet composite superfield
\begin{equation}
\mc{S}\mc{H}=\textrm{Tr}(\Theta^+_\Sigma\Theta^0_\Sigma\Theta^-_\Sigma)
\end{equation}
where $\Theta_\Sigma$ is the field strength superfield. The
components $(\Theta^+_\Sigma,\Theta^0_\Sigma,\Theta^-_\Sigma)$ are
denoted in the $N=2$ langauge by $(Y,\Sigma,-Y^\dagger)$ together
with derivative terms. The explicit form of these can be found in
\cite{Ring_N3_superfield}. Upon expanding in powers of the
superspace coordinates $(\theta^\pm, \theta^0)$, we obtain the
composite operators dual to the various component fields within the
massive gravitino multiplet. For example, the scalar operator of
dimension $6$ corresponding to the breathing mode of the $N^{010}$
manifold is given by the $N=3$ supersymmetrization of the operator
\begin{equation}
\epsilon^{\lambda\mu\nu}\epsilon^{\rho\sigma\tau}F_{\lambda\mu}F_{\nu\rho}F_{\sigma\tau}\,
.
\end{equation}
It should be noted that this operator is the highest component of
the supermultiplet with six factors of the $(\theta^{\pm},\theta^0)$
coordinates. The deformation corresponding to this operator is then
expected to preserve supersymmetry.
\\
\indent It has been pointed out in \cite{N010_truncation_Cassani}
that the SCFT dual to the $N=1$ critical point on the other hand
should be identified with the $N=1$ SCFT arising from the squashed
seven-sphere given in \cite{N1_squashed}. This is due to the similar
spectrum within the truncation of \cite{N010_truncation_Cassani} and
that of the squashed seven-sphere. However, very little is known
about $N=1$ SCFT in three dimensions apart from holographic
descriptions.

\section{$N=3$ supersymmetric solutions}\label{N3_solution}
We now look at the resulting BPS equations and their solutions. By
using the coset representative \eqref{SO3_inv_coset}, we find that
$A^{ij}_1$ tensor is diagonal
\begin{equation}
A^{ij}_1=\textrm{diag}(\alpha_1,\alpha_3,\alpha_3,\alpha_3).
\end{equation}
The two eigenvalues $\alpha_1$ and $\alpha_3$ correspond to Killing spinors $\epsilon^1$ and $\epsilon^{2,3,4}$ and give rise to the superpotentials
\begin{eqnarray}
\mc{W}_1&=&\frac{3}{2}e^{-\frac{3}{2}(2U_1+V_1)}\left[e^{2U_1}+2e^{U_1+3V_1}+k-2\chi Z-Z^2 \right.\nonumber \\
& &\left. +2i\left[e^{U_1}\chi+(e^{U_1}+e^{3V_1})Z\right]\right],\\
\mc{W}_3&=&-\frac{1}{2}e^{-\frac{3}{2}(2U_1+V_1)}\left[5e^{2U_1}+2e^{U_1+3V_1}-3k+6\chi Z+3Z^2\right.\nonumber \\
& &\left.+2i\left[e^{U_1}(\chi+Z)-3e^{3V_1}Z\right]\right].
\end{eqnarray}
In this section, we will consider only $\mc{W}_3$ corresponding to unbroken $N=3$ supersymmetry and leave the analysis of $\mc{W}_1$ to the next section.

\subsection{Flow to $N=3$ non-conformal field theory}
The analysis of $\delta \lambda^i_a=0$ equations along
$\epsilon^{2,3,4}$ requires $U_1=3V_1$ and $\chi=2Z$. However, we
need to further set $\chi=Z=0$ in the BPS equations in order to
satisfy the field equations. With all these requirements, we end up
with the $N=3$ BPS equations
\begin{eqnarray}
V_1'&=&e^{-\frac{21}{2}V_1}(e^{6V_1}-k),\label{N3_eq1}\\
A'&=&\frac{1}{2}e^{-\frac{21}{2}V_1}(7e^{6V_1}-3k).\label{N3_eq2}
\end{eqnarray}
From these equations, we find an $N=3$ $AdS_4$ critical point
\begin{equation}
V_1=\frac{1}{6}\ln k,\qquad
A'=\frac{2}{k^{\frac{3}{4}}}=\frac{1}{L_3}\, .
\end{equation}
We also see that there is no critical point for $k<0$. This is in
agreement with the fact that the solutions with $k<0$ break all
supersymmetry as mentioned before. In equations \eqref{N3_eq1} and
\eqref{N3_eq2}, we have chosen a definite sign choice to obtain the
correct behavior near the critical point
\begin{equation}
V_1\sim e^{\frac{3r}{L_3}}\, .
\end{equation}
This is consistent with the fact that $V_1$ is dual to an irrelevant
operator of dimension six. We then see that the dual $N=3$ SCFT
appears in the IR.
\\
\indent It can be checked that these equations satisfy the scalar
field equations and Einstein equations. In this case, the
superpotential is real
\begin{equation}
\mc{W}_3=W_3=\frac{1}{2}e^{-\frac{21}{2}V_1}(7e^{6V_1}-3k),
\end{equation}
and the scalar potential can be written as
\begin{eqnarray}
V&=&\frac{4}{189}\left(\frac{\pd W_3}{\pd V_1}\right)-3W_3^2,\nonumber \\
&=&9k^2e^{-21V_1}-21e^{-9V_1}\, .\label{N3_potential}
\end{eqnarray}
For non-vanishing pseudoscalars $\chi$ and $Z$ and $U_1\neq 3V_1$, $N=3$ supersymmetry
is broken, and the scalar potential cannot be written in term of the
real superpotential $W_3$. It should also be noted that the
vanishing of $\chi$ and $Z$ rules out any supersymmetric Janus
solutions since the corresponding BPS equations cannot be consistent
for finite $\ell$. This is similar to the results of
\cite{warner_Janus} and \cite{N3_Janus} in which pseudoscalars are
required for supersymmetric Janus solutions to exist.
\\
\indent We now return to a supersymmetric RG flow solution. The BPS equations given above have a simple solution
\begin{eqnarray}
A&=&\frac{3}{2}V_1+\frac{1}{3}\ln (e^{6V_1}-k),\\
V_1&=&-\frac{1}{6}\ln \left[\frac{1-e^{6k\tilde{r}+C}}{k}\right]
\end{eqnarray}
where the new radial coordinate $\tilde{r}$ is related to $r$ by
$\frac{d\tilde{r}}{dr}=e^{-\frac{21}{2}V_1}$. As $\tilde{r}\sim
r\rightarrow -\infty$, we find $V_1\sim e^{6k\tilde{r}}\sim
e^{\frac{3r}{L}}$. As usual in flows to non-conformal field
theories, there is a singularity at $\tilde{r}\sim -\frac{C}{6k}$
which gives $V_1\rightarrow \infty$. Near this singularity, we find
\begin{equation}
V_1\sim -\frac{1}{6}\ln(6k\tilde{r}+C)\qquad \textrm{and}\qquad
A\sim \frac{7}{2}V_1\sim -\frac{7}{12}\ln(6k\tilde{r}+C)\, .
\end{equation}
In this limit, the scalar potential vanishes. This implies that the
singularity is physical according to the criterion of
\cite{Gubser_singularity}.
\\
\indent We can also see this by looking at the eleven-dimensional
metric and considering the criterion of \cite{Maldacena_Nunez_nogo}.
In the present case, we have $U=V_1$ and
\begin{eqnarray}
ds^2_{11}&=&e^{-7V_1}ds^2_4+e^{2V_1}(ds^2(B_{\textrm{QK}})+\eta^I\eta^I)\nonumber \\
&=&dx^2_{1,2}+(6k\tilde{r}+C)^{-\frac{7}{3}}d\tilde{r}^2+(6k\tilde{r}+C)^{-\frac{1}{3}}\left[ds^2(B_{\textrm{QK}})+\eta^I\eta^I\right],\nonumber
\\
G_4&=&-6kdx^0\wedge dx^1\wedge dx^2 \wedge d\tilde{r}\, .
\end{eqnarray}
By changing to a new coordinate $R$ via the relation $dR=(6k\tilde{r}+C)^{-\frac{7}{6}}d\tilde{r}$, we can write the metric as
\begin{equation}
ds^2_{11}=dx^2_{1,2}+dR^2+(kR)^2\left[ds^2(B_{\textrm{QK}})+\eta^I\eta^I\right]
\end{equation}
Near the singularity, we then see that the metric component $g_{00}$ is
bounded, $g_{00}^{(11)}=-e^{2A-7V_1}\rightarrow -1$. Therefore, the
singularity is also physical by the criterion of
\cite{Maldacena_Nunez_nogo}. This solution should be identified with
the flow from $E^{1,2}\times HK$, $HK$ being a Hyper-Kahler
manifold, to $AdS_4\times N^{010}$ studied in \cite{F4_nunezAdS6} by
using another approach.
\\
\indent It should also be noted that when $k=0$, $AdS_4$
critical points do not exist. In this case, the gauged supergravity
however admits an $N=3$ supersymmetric domain wall vacuum. This
solution preserves only six supercharges due to the $\gamma_r$
projection and accordingly is a half-BPS solution. By setting $k=0$
in the BPS equations, we can find a simple domain wall solution
\begin{equation}
V_1=\frac{2}{9}\ln \frac{9r}{2},\qquad A=\frac{7}{9}\ln \frac{9r}{2}
\end{equation}
where, for convenience, we have set the associated integration
constants to zero by shifting the coordinates. This solution can be
readily lifted to eleven dimensions in which the metric is given by
\begin{eqnarray}
ds^2_{11}&=&dx^2_{1,2}+\left(\frac{9r}{2}\right)^{-\frac{14}{9}}dr^2+\left(\frac{9r}{2}\right)^{\frac{4}{9}}ds^2(B_{\textrm{QK}})+\eta^I\eta^I,\\
\nonumber
\\ &=& dx^2_{1,2}+dR^2+R^2\left(ds^2(B_{\textrm{QK}})+\eta^I\eta^I\right)
\end{eqnarray}
where we have defined a new coordinate
$R=\left(\frac{9r}{2}\right)^{\frac{2}{9}}$. In this case, the
four-form field vanishes.
\\
\indent As a final comment on the $N=3$ solution, we can also give a
geometric interpretation of the condition $U_1=3V_1$. Recall that
$U_1=3V_1$ means $U=V_1$, we find that only the breathing mode is
consistent with $N=3$ supersymmetry. As mentioned previously, the
breathing mode corresponds to an operator which is the highest
component of the supermultiplet and hence does not break
supersymmetry. On the other hand, the squashing mode corresponding
to the scalar $V_1-U$, dual to a dimension-$4$ operator, breaks all
of the supersymmetry. Non-supersymmetric RG flows between $N=(3,0)$
and $N=(0,1)$ supersymmetric $AdS_4$ critical points driven by this
scalar have been studied in \cite{N3_flow_Ahn2}, see also
\cite{N3_flow_Ahn1}. The dual operator driving the flow has also
been proposed in \cite{N3_flow_Ahn2}.

\section{$N=1$ supersymmetric solutions}\label{N1_solution}
In this section, we will carry out a similar analysis for the case
of unbroken $N=1$ supersymmetry corresponding to the Killing spinor
$\epsilon^1$. The real superpotential is given by
\begin{equation}
W_1=\frac{3}{2}e^{-3U_1-\frac{3}{2}V_1}\sqrt{[2\chi
e^{U_1}+2Z(e^{U_1} +e^{3V_1})]^2+[e^{2U_1}+2e^{U_1+3V_1}+k-2\chi Z
-Z^2]^2}\label{N1_superpotential}
\end{equation}
in term of which the scalar potential can be written as
\begin{equation}
V=-2G^{\alpha\beta}\frac{\pd W_1}{\pd \phi^\alpha}\frac{\pd W_1}{\pd \phi^\beta}-3W_1^2
\end{equation}
where $\phi^\alpha=(U_1,V_1,Z,\chi)$ and $G^{\alpha\beta}$ is the
inverse of the metric in the scalar kinetic terms given in
\eqref{scalar_kin}. We now look at the BPS equations and possible
supersymmetric solutions.

\subsection{RG flow solutions}
\begin{figure}
         \centering
         \begin{subfigure}[b]{0.3\textwidth}
                 \includegraphics[width=\textwidth]{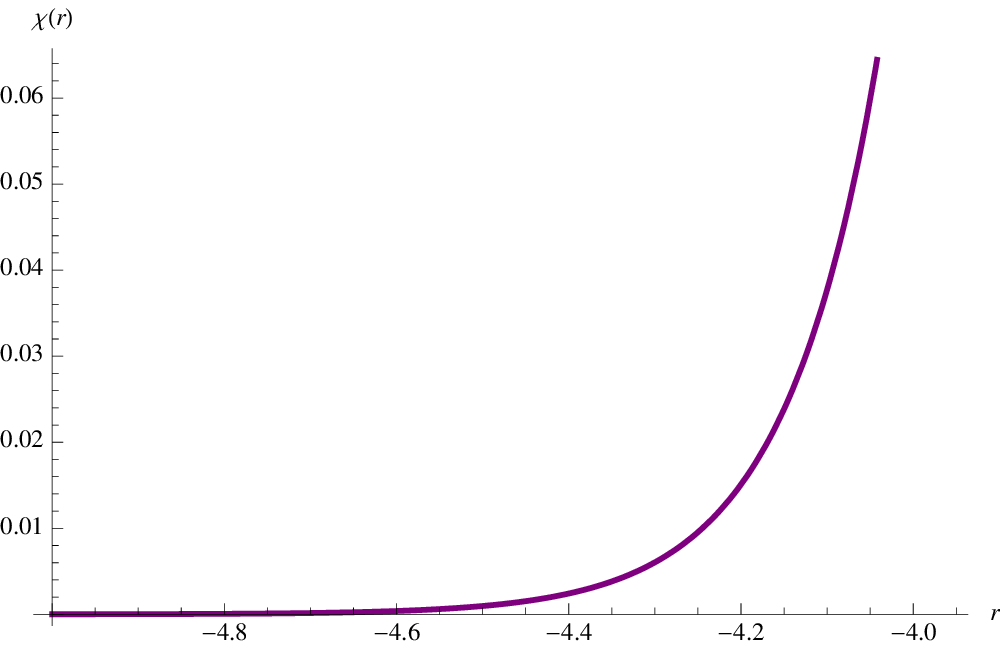}
                 \caption{Solution for $\chi$}
         \end{subfigure}%
         ~ 
         \begin{subfigure}[b]{0.3\textwidth}
                 \includegraphics[width=\textwidth]{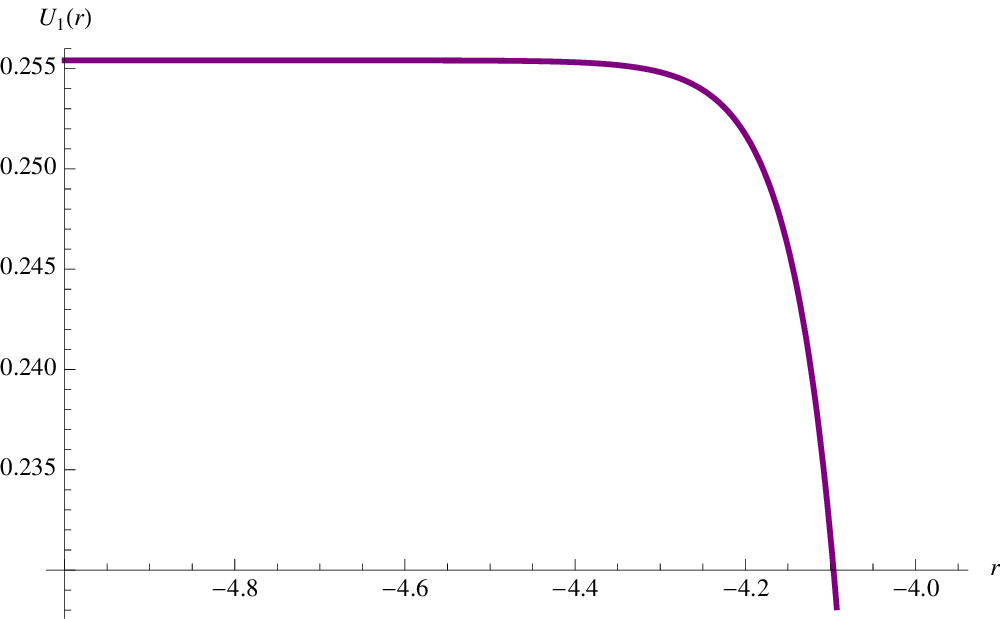}
                 \caption{Solution for $U_1$}
         \end{subfigure}
         \begin{subfigure}[b]{0.3\textwidth}
                 \includegraphics[width=\textwidth]{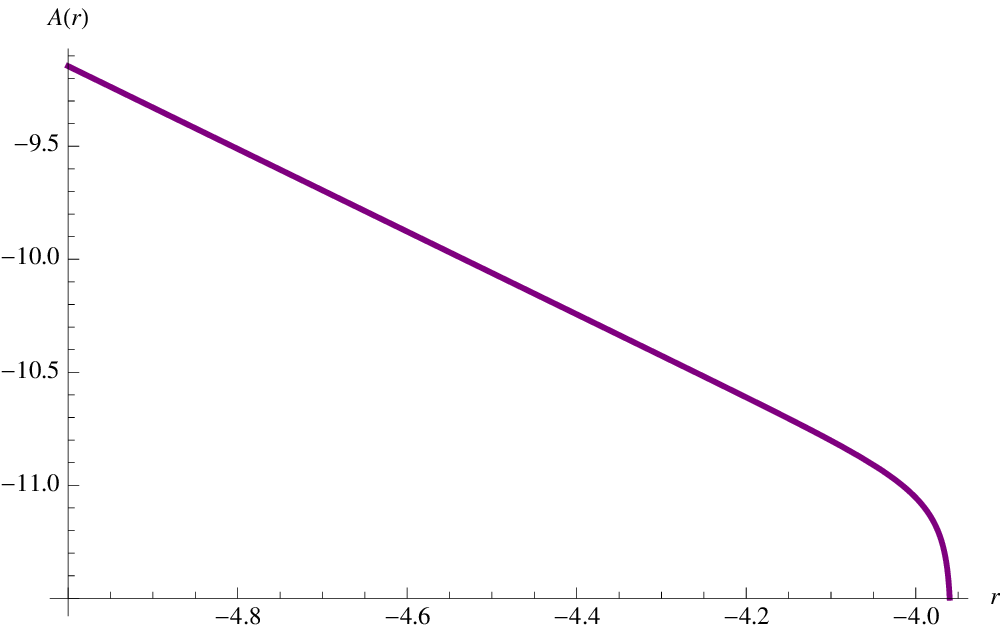}
                 \caption{Solution for $A$}
         \end{subfigure}\\
\begin{subfigure}[b]{0.3\textwidth}
                 \includegraphics[width=\textwidth]{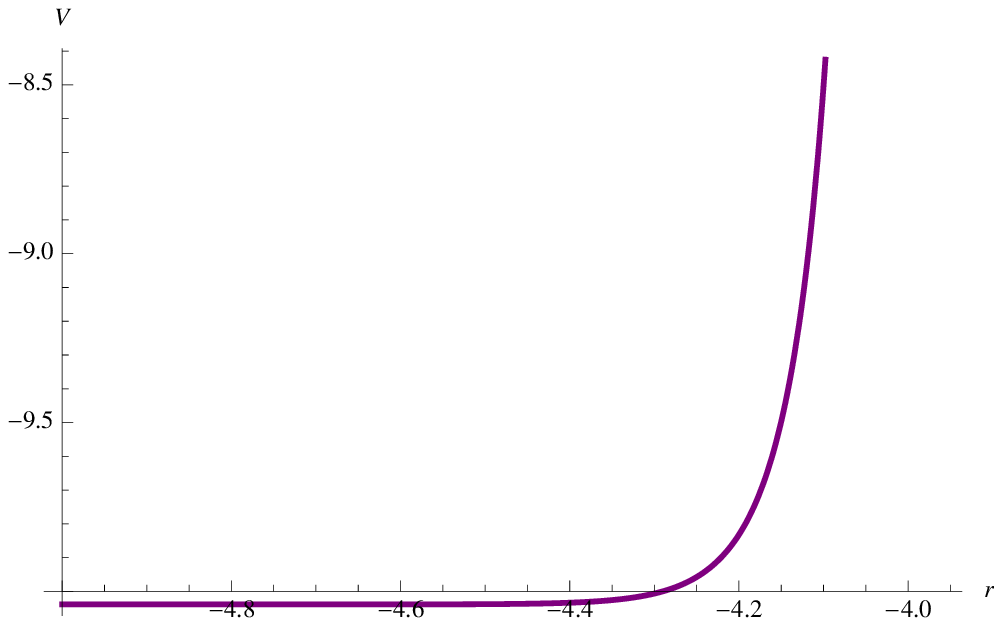}
                 \caption{The scalar potential along the flow}
         \end{subfigure}\quad%
         ~ 
         \begin{subfigure}[b]{0.3\textwidth}
                 \includegraphics[width=\textwidth]{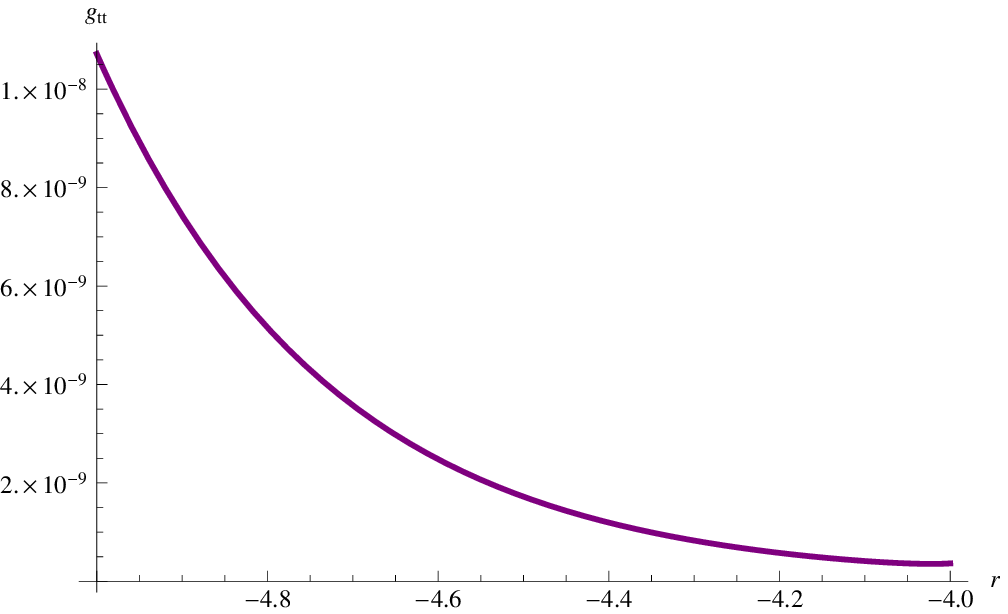}
                 \caption{The metric component $g^{(11)}_{00}$}
         \end{subfigure}
         \caption{An RG flow solution from $N=1$ non-conformal field theory to $N=1$ SCFT with one scalar and one pseudoscalar and $k=-1$}\label{fig1}
 \end{figure}

\begin{figure}
         \centering
         \begin{subfigure}[b]{0.3\textwidth}
                 \includegraphics[width=\textwidth]{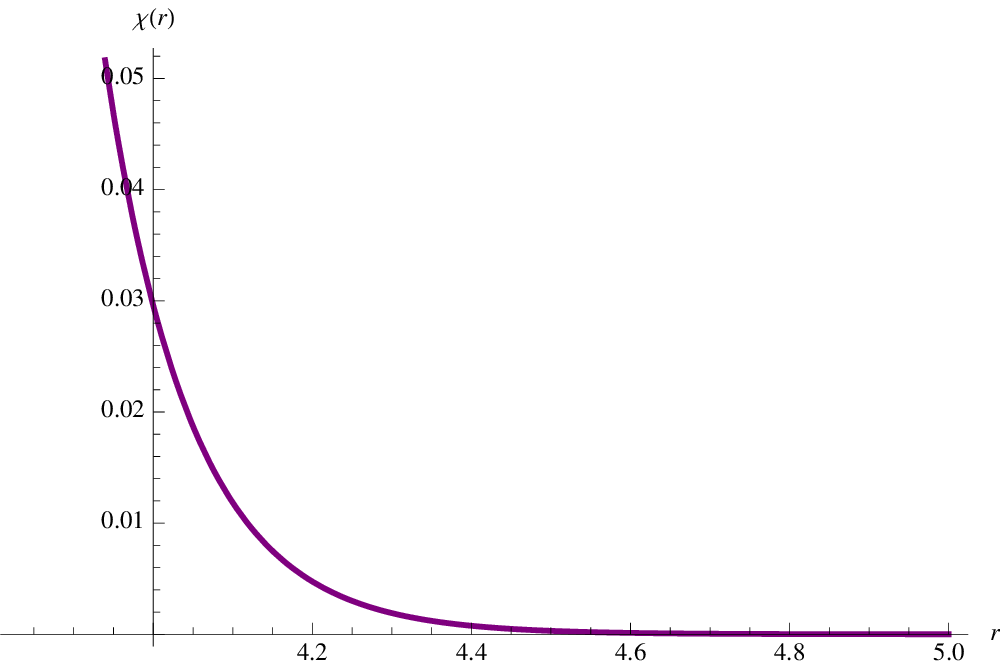}
                 \caption{Solution for $\chi$}
         \end{subfigure}%
         ~ 
         \begin{subfigure}[b]{0.3\textwidth}
                 \includegraphics[width=\textwidth]{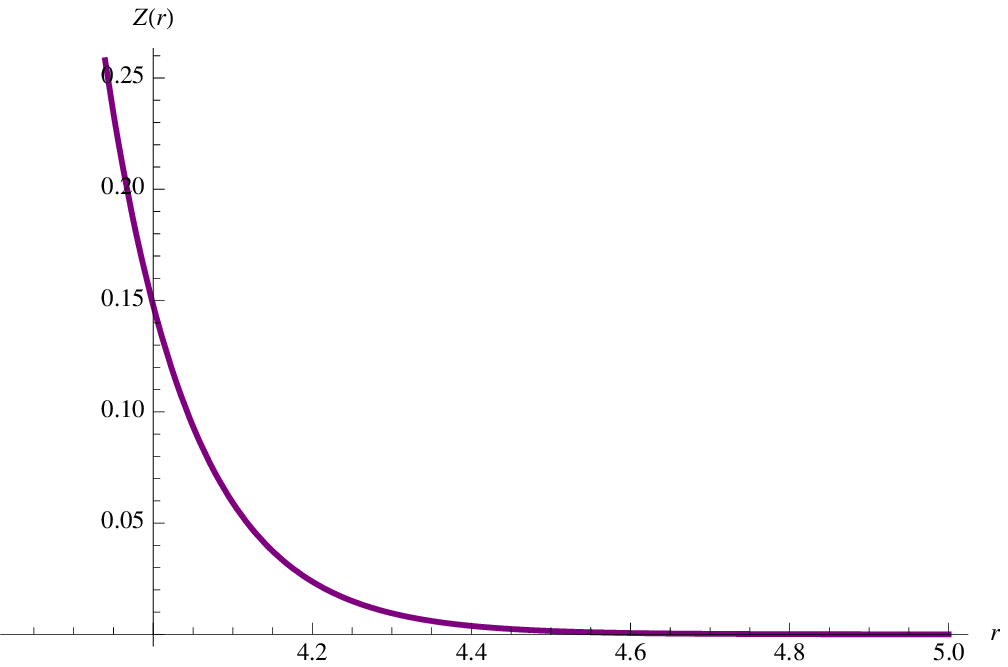}
                 \caption{Solution for $Z$}
         \end{subfigure}
         \begin{subfigure}[b]{0.3\textwidth}
                 \includegraphics[width=\textwidth]{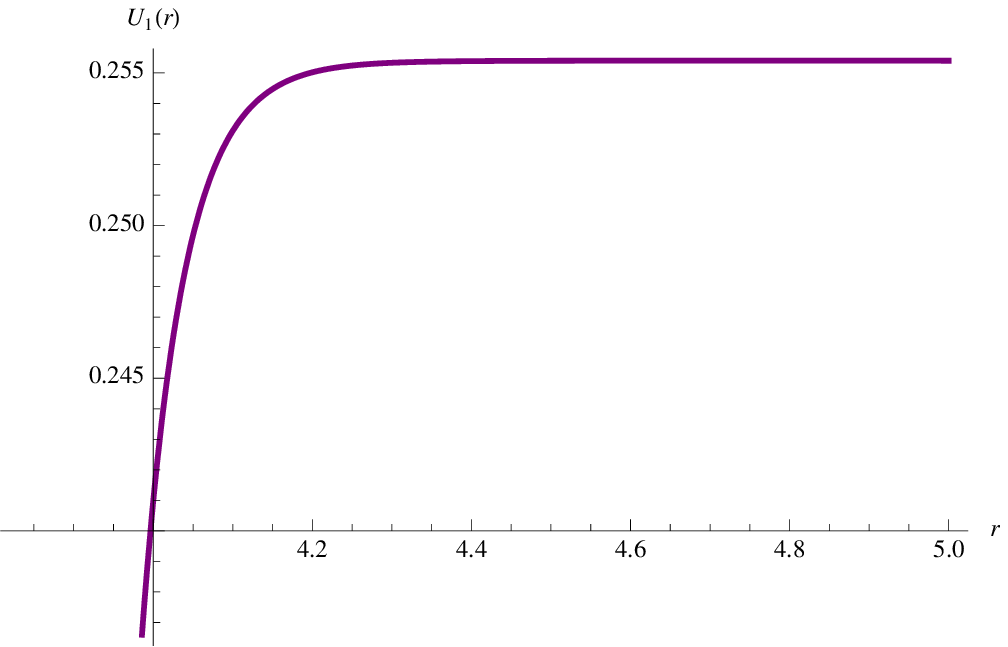}
                 \caption{Solution for $U_1$}
         \end{subfigure}\\
         \begin{subfigure}[b]{0.3\textwidth}
                 \includegraphics[width=\textwidth]{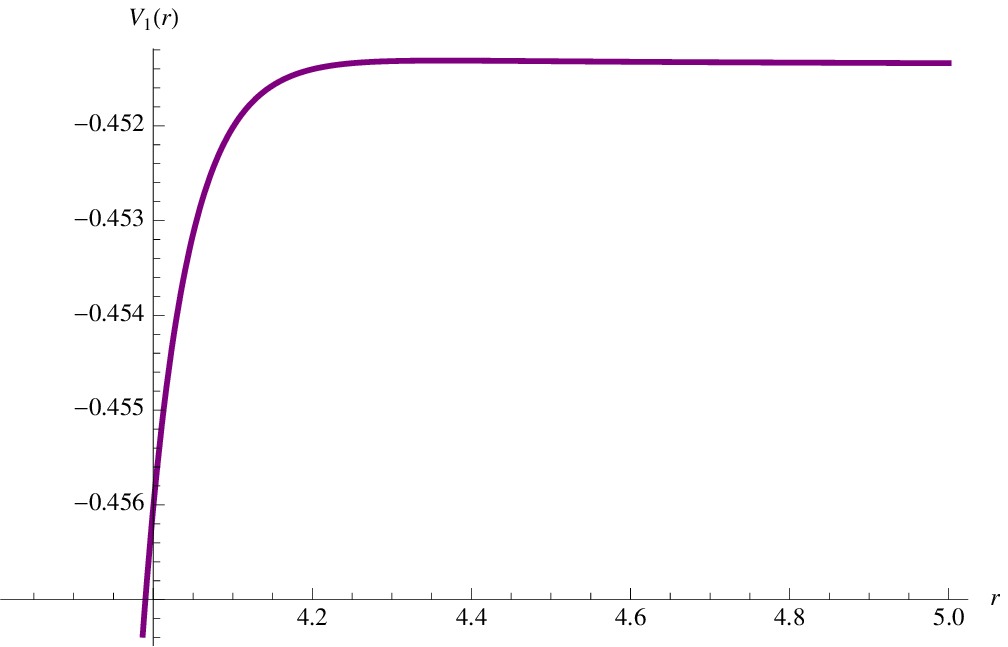}
                 \caption{Solution for $V_1$}
         \end{subfigure}\quad%
         ~ 
         \begin{subfigure}[b]{0.3\textwidth}
                 \includegraphics[width=\textwidth]{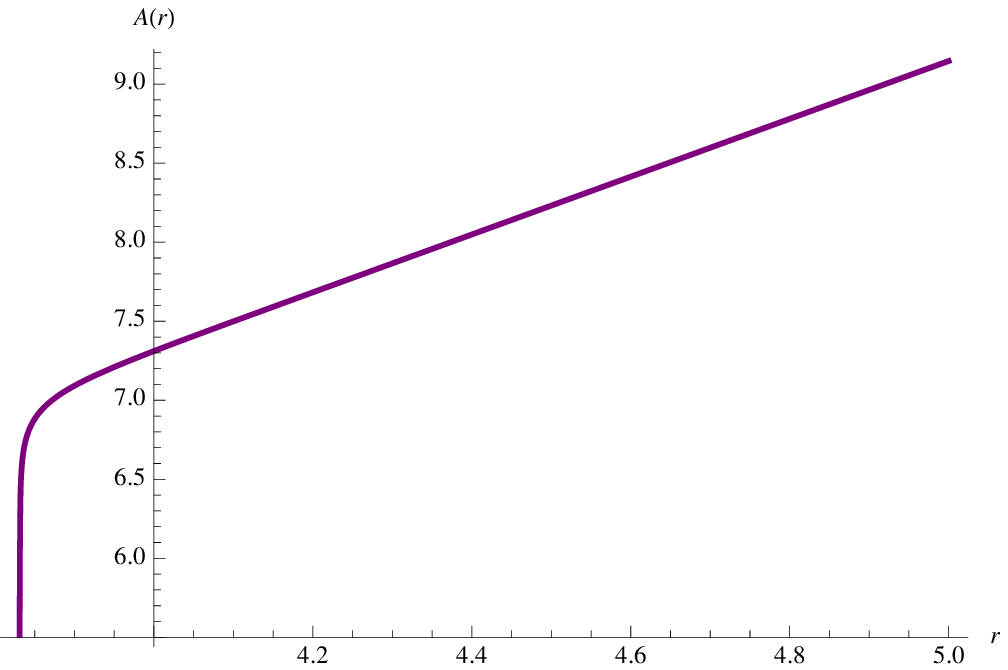}
                 \caption{Solution for $A$}
         \end{subfigure}\\
         \begin{subfigure}[b]{0.3\textwidth}
                 \includegraphics[width=\textwidth]{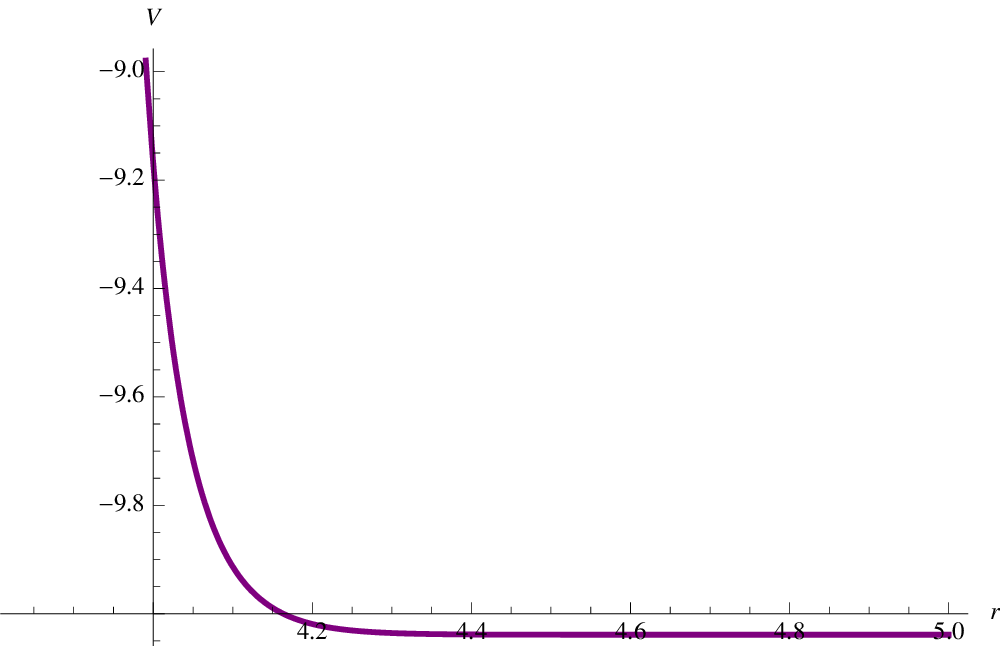}
                 \caption{The scalar potential along the flow}
         \end{subfigure}\quad%
         ~ 
         \begin{subfigure}[b]{0.3\textwidth}
                 \includegraphics[width=\textwidth]{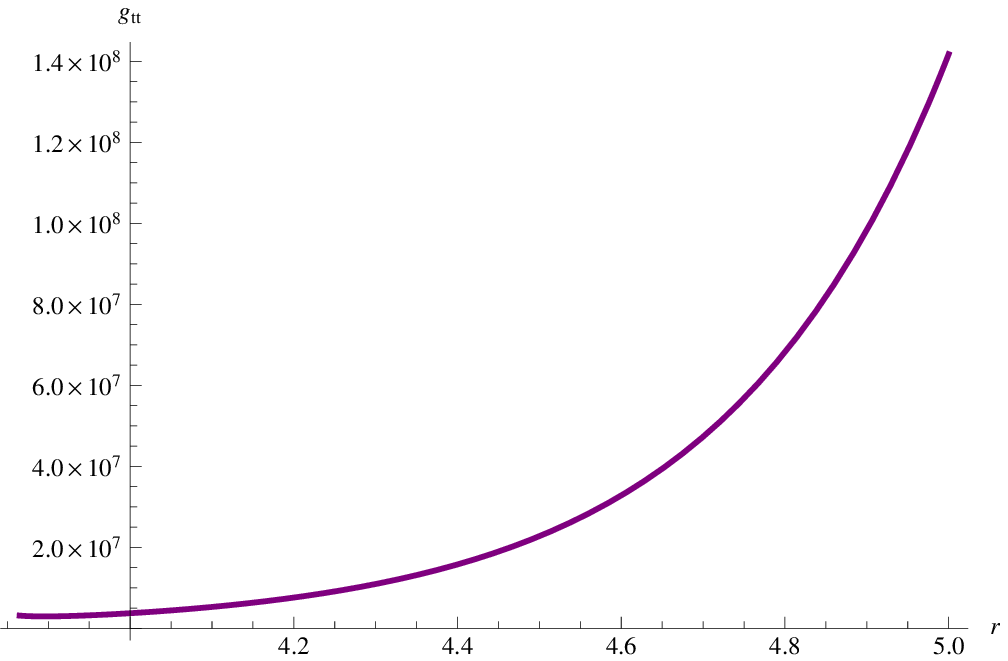}
                 \caption{The metric component $g^{(11)}_{00}$}
         \end{subfigure}
         \caption{An RG flow solution from $N=1$ SCFT to $N=1$ non-conformal field theory with two scalars and two pseudoscalars and $k=-1$}\label{fig2}
 \end{figure}

We begin with an RG flow solution with only $U_1$ and $V_1$ scalars
non-vanishing. These correspond to the breathing and squashing modes
of $N^{010}$. It can be checked that keeping only $U_1$ and $V_1$ is
consistent with the BPS equations and the corresponding field equations. From eleven-dimensional point of view, this
corresponds to pure metric modes since the pseudoscalars $Z$ and
$\chi$ appear in the internal components of the four-form field
strength. A non-supersymmetric flow between this $N=1$ $AdS_4$ and
the skew-whiffle $N=3$ $AdS_4$ has already been studied in
\cite{N3_flow_Ahn2} and \cite{N3_flow_Ahn1}.
\\
\indent In this work, we will study a supersymmetric flow to a
non-conformal field theory. The BPS equations in this case are given
by
\begin{eqnarray}
U_1'&=&e^{-\frac{3}{2}(2U_1+V_1)}(e^{2U_1}+4e^{U_1+3V_1}+3k),\\
V_1'&=&e^{-\frac{3}{2}(2U_1+V_1)}(e^{2U_1}-2e^{U_1+3V_1}+k),\\
A'&=&\frac{3}{2}e^{-\frac{3}{2}(2U_1+V_1)}(e^{2U_1}+2e^{U_1+3V_1}+k).
\end{eqnarray}
From these equations, we clearly see that there is only one $AdS_4$
critical point given by the $N=1$ critical point II in section
\ref{N4theory}, and there exists a critical point only for $k<0$ as
previously remarked.
\\
\indent Near this $N=1$ critical point, we find an asymptotic
behavior
\begin{equation}
3V_1-U_1\sim e^{-\frac{5r}{3L}},\qquad 2U_1+V_1\sim e^{\frac{3r}{L}}
\end{equation}
corresponding to relevant and irrelevant operators of dimensions
$\Delta=\frac{5}{3},\frac{4}{3}$ and $\Delta=6$, respectively.
\\
\indent We begin with a simple case in which the relevant
deformation is further truncated out. This can be achieved by
setting $V_1=\frac{U_1}{3}-\frac{1}{3}\ln 5$. By taking appropriate
combinations, we find new BPS equations
\begin{eqnarray}
3V_1'-U'_1&=&2e^{-2U_1-\frac{3}{2}V_1}(e^{U_1}-5e^{3V_1}),\label{N1_eq11}\\
2U_1'+V_1'&=&e^{-\frac{3}{2}V_1-3U_1}(3e^{2U_1}+6e^{U_1+3V_1}+7k)\label{N1_eq12}
\end{eqnarray}
from which we immediately see that the above truncation is
consistent. Under this truncation, the remaining BPS equations become
\begin{eqnarray}
U_1'&=&\frac{3}{\sqrt{5}}e^{-\frac{7}{2}U_1}\left(3e^{2U_1}+5k\right),\\
A'&=&\frac{3}{2\sqrt{5}}e^{-\frac{7}{2}U_1}\left(7e^{2U_1}+5k\right).
\end{eqnarray}
By changing to a new radial coordinate $\tilde{r}$, defined by
$\frac{d\tilde{r}}{dr}=e^{-\frac{7}{2}U_1}$, as in the $N=3$ case,
we obtain a solution
\begin{eqnarray}
U_1&=&-\frac{1}{2}\ln
\left[\frac{e^{-6\sqrt{5}k\tilde{r}}-3}{5k}\right],\nonumber \\
A&=&\frac{1}{2}U_1+\frac{1}{3}\ln \left(6e^{2U_1}+10k\right)
\end{eqnarray} where we have absorbed all
integration constants by shifting $\tilde{r}$ and rescaling
$dx^2_{1,2}$ coordinates. It should also be remembered that in this
case $k<0$. The singularity at $6\sqrt{5}k\tilde{r}\rightarrow
-\ln3$ is physical by the criterions of both
\cite{Gubser_singularity} and \cite{Maldacena_Nunez_nogo}. In this
case, we find, as $6\sqrt{5}k\tilde{r}\rightarrow -\ln3$,
\begin{equation}
V\rightarrow 0,\qquad g_{00}^{(11)}\rightarrow -5^{\frac{1}{3}}\, .
\end{equation}
We identify this solution with the flow from $E^{1,2}\times Spin(7)$
to $AdS_4\times \tilde{S}^7$ where $\tilde{S}^7$ is the squashed
seven-sphere with a weak $G_2$ holonomy.
\\
\indent To solve equations \eqref{N1_eq11} and \eqref{N1_eq12}
in the presence of both types of deformations, we introduce new
scalar fields defined by
\begin{equation}
\tilde{V}=3V_1-U_1\qquad \textrm{and}\qquad \tilde{U}=2U_1+V_1
\end{equation}
in terms of which the BPS equations become
\begin{eqnarray}
\tilde{V}'&=&e^{-\frac{2}{7}\tilde{V}-\frac{9}{14}\tilde{U}}(2-10e^{\tilde{V}}),\label{Vt_eq}\\
\tilde{U}'&=&e^{-\frac{3}{2}\tilde{U}}\left(3e^{-\frac{2}{7}(\tilde{V}-2\tilde{U})}
+6e^{\frac{5}{7}\tilde{V}+\frac{6}{7}\tilde{U}}+7k\right),\label{Ut_eq}\\
A'&=&\frac{3}{2}e^{-\frac{2}{7}\tilde{V}-\frac{3}{2}\tilde{U}}\left(2e^{\tilde{V}+\frac{6}{7}\tilde{U}}
+e^{\frac{6}{7}\tilde{U}}+ke^{\frac{2}{7}\tilde{V}}\right).\label{At_eq}
\end{eqnarray}
We then define a new coordinate $\rho$ via the relation
\begin{equation}
\frac{d\rho}{dr}=e^{-\frac{2}{7}\tilde{V}-\frac{9}{14}\tilde{U}}\, .
\end{equation}
An analytic solution to the above equations can subsequently be
obtained
\begin{eqnarray}
\tilde{V}&=&-\ln \left(5+C_1e^{2\rho}\right),\nonumber \\
\tilde{U}&=&\frac{7}{3}\rho
+\frac{7}{6}\ln\left[-3k(5+C_1e^{-2\rho})^{\frac{5}{7}}\right.
\nonumber
\\
& &\left.(5+C_1e^{-2\rho})^{\frac{18}{35}}\left[3k(5)^{\frac{1}{5}}
\phantom{}_2F_1\left(\frac{4}{5},\frac{4}{5},\frac{9}{5},-\frac{C_1}{5}e^{-2\rho}\right)
-\frac{162}{7}(3^{\frac{3}{5}})C_1C_2e^{\frac{8}{5}\rho}\right]
\right],\nonumber \\
A&=&\frac{6}{7}\rho+\frac{3}{14}\tilde{U}+\frac{6}{35}\ln
\left[8C_1+40e^{2\rho}\right]
\end{eqnarray}
where $_2F_1$ is the hypergeometric function.
\\
\indent We now consider the asymptotic behavior of this RG flow to
$N=1$ non-conformal field theories. Near the singularity at $\rho
=\frac{1}{2}\ln \left(-\frac{C_1}{5}\right)$, we find that
\begin{eqnarray}
\tilde{V}&=&
-\ln\left(5+C_1e^{-2\rho}\right)\rightarrow\infty,\nonumber \\
\tilde{U}&\sim&\frac{3}{5}\ln\left(C_1+5e^{2\rho}\right)\rightarrow
-\infty,\nonumber\\
A&\sim&\frac{3}{10}\ln \left(C_1+5e^{2\rho}\right).
\end{eqnarray}
Although the scalar potential diverges near this singularity, the
eleven dimensional metric gives $g^{(11)}_{00}\sim
\textrm{constant}$. The singularity is then physical and the
solution describes an RG flow between the dual $N=1$ SCFT and a non-conformal
$N=1$ field theory. Near this singularity, the corresponding
eleven-dimensional solution is given by
\begin{eqnarray}
ds^2_{11}&=&dx^2_{1,2}+\frac{d\rho^2}{(C_1+5e^{2\rho})^{\frac{2}{5}}}
+(C_1+5e^{2\rho})^{\frac{3}{5}}ds^2(B_{\textrm{QK}})+(C_1+5e^{2\rho})^{-\frac{2}{5}}\eta^I\eta^I,\nonumber \\
& &\\
G_4&=&-6k(C_1+5e^{2\rho})^{-\frac{8}{35}}dx^0\wedge dx^1\wedge
dx^2\wedge \rho
\end{eqnarray}
where we have absorbed a constant in the $dx^2_{1,2}$ coordinates.
\\
\indent We then move to more complicated RG flows involving the $SO(3)_R$ singlet pseudoscalars. In this case, the flows will
involve the internal components of the four-form field strength.
Before considering possible solutions, we give an explicit
form of the uplift formulae for the metric and the four-form with
non-vanishing $Z$ and $\chi$
\begin{eqnarray}
ds^2_{11}&=&e^{-(2U_1+V_1)}ds^2_4+e^{U_1-V_1}ds^2(B_{\textrm{QK}})+e^{2V_1}\eta^I\eta^I,\nonumber
\\
G_4&=&-6ke^{-6U_1-3V_1}\textrm{vol}_4+12Z\textrm{vol}(\textrm{QK})+\chi'dr\wedge
\eta^1\wedge \eta^2 \wedge \eta^3\nonumber \\
& &+(\chi+Z)\epsilon_{IJK}\eta^I\wedge \eta^J\wedge J^K+Z'dr\wedge
\eta^I \wedge J^I\, .\label{N1_uplift}
\end{eqnarray}
\indent The $N=1$ BPS equations with four non-vanishing scalars are
given by
\begin{eqnarray}
U_1'&=&-\frac{2}{3}\frac{\pd W_1}{\pd U_1},\qquad
V_1'=-\frac{4}{9}\frac{\pd W_1}{\pd V_1},\nonumber \\
Z'&=&-\frac{2}{3}e^{2U_1}\frac{\pd W_1}{\pd Z},\qquad
\chi'=-4e^{6V_1}\frac{\pd W_1}{\pd \chi},\qquad A'=W_1
\end{eqnarray}
where the superpotential $W_1$ is given in
\eqref{N1_superpotential}. The explicit form of these equations can
be found in appendix \ref{N1_BPS_eq}.
\\
\indent We will begin with the solutions near the $N=1$ $AdS_4$
critical point. Near this critical point with $r\rightarrow \infty$, we find that
\begin{eqnarray}
3V_1-U_1&\sim& e^{-\frac{5r}{3L_1}},\qquad 2U_1+V_1\sim e^{\frac{3r}{L_1}},\nonumber \\
\chi+\frac{6}{5}Z&\sim& e^{-\frac{5r}{L_1}},\qquad
\chi-\frac{Z}{5}\sim e^{-\frac{r}{3L_1}}\, .\label{N1_asymptotic}
\end{eqnarray}
From these, we see that $U_1$ and $V_1$ are combinations of a
relevant and an irrelevant operators of dimensions
$\Delta=\frac{5}{3},\frac{4}{3}$ and $\Delta=6$ as in the previous
case while $Z$ and $\chi$ are combinations of a relevant and an
irrelevant operators of dimensions $\Delta =\frac{8}{3}$ and
$\Delta=5$, respectively. These are consistent with the scalar
masses given in \cite{N010_truncation_Cassani}.
\\
\indent Even with pseudoscalars turned on, there is a consistent
truncation keeping only irrelevant scalars. This truncation is given
by
\begin{equation}
V_1=\frac{1}{3}U_1-\frac{1}{3}\ln 5\qquad \textrm{and} \qquad
Z=5\chi\, .
\end{equation}
Within this truncation, the BPS equations become
\begin{eqnarray}
U_1'&=&\frac{e^{-\frac{7}{2}U_1}}{\sqrt{5}\tilde{W}}\left[63e^{4U_1}+150ke^{2U_1}+75k^2
+5250\chi^2(e^{2U_1}-k)+91875\chi^4\right],\label{U1_eq2}\quad \\
\chi'&=&-\frac{12\chi
e^{-\frac{3}{2}U_1}\left(7e^{2U_1}-5k+175\chi^2\right)}{\sqrt{5}\tilde{W}},\label{Chi_eq2}\\
A'&=&\frac{3e^{-\frac{7}{2}U_1}}{2\sqrt{5}}\tilde{W}\label{A_eq2}
\end{eqnarray}
where
\begin{equation}
\tilde{W}=\sqrt{(7e^{2U_1}+5k)^2+350\chi^2(7e^{2U_1}-5k)+30625\chi^4}\, .
\end{equation}
In this case, the BPS equations cannot be completely solved
analytically. However, the solution can be implicitly given by
defining a new scalar field $F$ via the relation $F=e^{2U_1}$ in
term of which the BPS equations read
\begin{eqnarray}
\frac{d\chi}{dr}&=&-\frac{12\chi\left(175\chi^2+7F-5k\right)}
{\sqrt{5}F^{\frac{3}{4}}\sqrt{49F^2+25(k-35\chi^2)^2+70F(k+35\chi^2)}},\label{Chi_r_eq}\\
\frac{dF}{d\chi}&=&-\frac{21F^2+50F(k+35\chi^2)+25(k-35\chi^2)^2}{2\chi\left(7F+175\chi^2-5k\right)},\label{F_chi_eq}\\
\frac{dA}{d\chi}&=&-\frac{49F^2+70F(k+35\chi^2)+25(k-35\chi^2)^2}{8\chi
F\left(7F+175\chi^2-5k\right)}\label{A_chi_eq}
\end{eqnarray}
where in the last two equations we have taken $\chi$ as an
independent variable by combing $F'$ and $A'$ equations with $\chi'$
equation, respectively. By solving equation \eqref{F_chi_eq}, we can
determine $F(\chi)$ implicitly from the following solution
\begin{eqnarray}
C\chi&=&2^{\frac{2}{5}}7^{\frac{1}{5}}\left[5(k35\chi^2)+7F\right]\phantom{}_2F_1\left(\frac{1}{2},\frac{4}{5},
\frac{3}{2},\frac{\left[5(k+35\chi^2)+7F\right]^2}{3500k\chi^2}\right)\nonumber
\\
&
&-175(5^{\frac{2}{5}})\chi^2\left[\frac{49F^2+70F(k+35\chi^2)+25(k-35\chi^2)^2}{k\chi^2}\right]^{\frac{1}{5}}\,
.
\end{eqnarray}
In principle, $F(\chi)$ can be substituted in equations
\eqref{Chi_r_eq} and \eqref{A_chi_eq} to determine $\chi(r)$ and $A(\chi)$.
\\
\indent We now look for asymptotic behavior for large values of
scalar fields. At large $\chi$, we find that
\begin{equation}
U_1=\frac{1}{2}\ln F\sim\frac{1}{2}\ln
\frac{C}{\chi^\frac{3}{2}},\qquad \chi\sim C'r^{-\frac{8}{9}}
\end{equation}
where for convenience we have shifted the coordinate $r$ such that
the singularity is present at $r=0$. In genral, $\chi(r\rightarrow 0)$ can be $\infty$ or $-\infty$ depending on the sign of the constant $C'$. For definiteness, we will take $C'>0$ in the present discussion. This behavior give the metric warped factor
\begin{equation}
A\sim \frac{7}{9}\ln r\, .
\end{equation}
Near the singularity, we find that the scalar potential diverges but
$g^{(11)}_{00}$ becomes constant. We then conclude that the
singularity is physical by the criterion of
\cite{Maldacena_Nunez_nogo}. For completeness, we give an example of
numerical solutions with $k=-1$ in figure \ref{fig1}.
\\
\indent Note that we have identified the $N=1$ $AdS_4$ critical point at
$\chi=0$ and $U_1=0.22541$, for $k=-1$, with the IR SCFT at
$r=-\infty$. The numerical solution also gives a singularity
consistent with the above analysis namely the divergence of scalars
and the potential as well as the constancy of $g_{00}^{(11)}$. The eleven-dimensional solution near the singularity can be obtained as follow
\begin{eqnarray}
ds^2_{11}&=&dx^2_{1,2}+dR^2+\left(\frac{2R}{3}\right)^2\left[ds^2(B_{\textrm{QK}})+\frac{1}{5}\eta^I\eta^I\right],\nonumber \\
G_4&=&-6k\left(\frac{2R}{9}\right)^{\frac{7}{2}}dx^0\wedge dx^1\wedge dx^2\wedge dr+60\left(\frac{2R}{9}\right)^{-4}\textrm{vol}(B_{\textrm{QK}})\nonumber \\
& &+6\left(\frac{2R}{9}\right)^{-4}\epsilon_{IJK}\eta^I\wedge \eta^J\wedge J^K-\left(\frac{2}{9}\right)^{-\frac{17}{2}}R^{-5}dR\wedge \eta^1\wedge \eta^2\wedge \eta^3\nonumber \\
& &-5\left(\frac{2}{9}\right)^{-\frac{17}{2}}R^{-5}dR\wedge \eta^I\wedge J^I
\end{eqnarray}
where we have defined a new coordinate $R=\frac{9}{2}r^{\frac{2}{9}}$.
\\
\indent We now look at the most general flow solution with all four
scalars turned on. The BPS equations are too complicated to be
solved analytically. In any case, numerical solutions can be
obtained by suitable boundary conditions similar to the previous
case. From the asymptotic behaviour of these scalars given in \eqref{N1_asymptotic}, there
could be many possible singularities at the end of the flows due to
the presence of various vacuum expectation values (vev) and operator
deformations as in the solutions studied in \cite{Yi_4D_flow}. We
will only give an example of these solutions. This is shown in
figure \ref{fig2} in which we take $k=-1$, and the $N=1$ critical point
corresponds to the values of the scalar fields
\begin{equation}
U_1=0.25541,\qquad V_1=-0.45134,\qquad Z=\chi=0\, .
\end{equation}
\indent Near the singularity, we can see that $\chi\sim Z\rightarrow
\infty$ and $U_1\sim V_1\rightarrow -\infty$. From the BPS
equations, we can make an analysis near this limit resulting in the
asymptotic behavior
\begin{eqnarray}
U_1\sim 3V_1\sim \frac{1}{2}\ln \chi^{-\frac{3}{2}},\qquad Z\sim
5\chi\sim r^{-\frac{8}{9}}\, .
\end{eqnarray}
Using these expressions or the numerical analysis in figure \ref{fig2}, we can
see that the singularity is physical due to the constancy of
$g^{(11)}_{00}$ although the scalar potential becomes infinite. The uplift of this solution can be obtained along the same line as in the previous case. 

\subsection{Domain wall solutions}
Similar to the $N=3$ case, we will consider $N=1$ domain wall
solutions to the BPS equations with $k=0$. All of the relevant BPS
equations can be obtained from those given above by setting $k=0$,
so we will not repeat them here.
\\
\indent In the case of vanishing pseudoscalars, we find a domain
wall solution to equations \eqref{Vt_eq}, \eqref{Ut_eq} and
\eqref{At_eq} with $k=0$
\begin{eqnarray}
\tilde{U}&=&\frac{3}{2}\tilde{V}-\frac{21}{20}\ln\left[105e^{\tilde{V}}-21\right],\qquad
A=\frac{1}{2}\tilde{U},\nonumber \\
r&=&\frac{2e^{\frac{1}{4}\tilde{V}}\left[2+25e^{\tilde{V}}-2\left(1-5e^{\tilde{V}}\right)^{\frac{27}{20}}
\phantom{}_2F_1\left(\frac{1}{4},\frac{7}{20},\frac{5}{4},5e^{\tilde{V}}\right)\right]}
{3969(21)^{\frac{7}{20}}\left(5e^{\tilde{V}}-1\right)^{\frac{27}{20}}}\,
.
\end{eqnarray}
The last equation implicitly gives the scalar $\tilde{V}(r)$.
\\
\indent With non-vanishing pseudoscalars, we find an analytic
solution only for the subtruncation to irrelevant scalars,
$V_1=\frac{1}{3}U_1-\frac{1}{3}\ln 5$ and $Z=5\chi$. The solution to
equations \eqref{U1_eq2}, \eqref{Chi_eq2} and \eqref{A_eq2} with
$k=0$ is given by
\begin{eqnarray}
U_1&=&\frac{1}{2}\ln
\left[\frac{C_1}{2\chi^{\frac{3}{2}}}-25\chi^2\right],\nonumber \\
A&=&\frac{1}{4}\left(50\chi^{\frac{7}{2}}-C_1\right)-\frac{7}{8}\ln\chi,\nonumber
\\
r&=&\frac{2\sqrt{5}2^{\frac{1}{4}}(C_1-50\chi^{\frac{7}{2}})+525\chi^{\frac{13}{8}}
\left(50\chi^{\frac{7}{2}}-C_1\right)^{\frac{1}{4}}\phantom{}_2F_1\left(-\frac{1}{7},\frac{1}{4},\frac{6}{7},
\frac{C_1}{50\chi^{\frac{7}{2}}}\right)}
{54\chi^{\frac{9}{8}}(C_1-50\chi^{\frac{7}{2}})^{\frac{1}{4}}}\, .\qquad
\end{eqnarray}
When uplifted to eleven dimensions, these solutions will provide domain walls with internal four-form fluxes. All of these solutions should
describe non-conformal $N=1$ field theories in three dimensions
according to the DW/QFT correspondence
\cite{DW/QFT_townsend,correlator_DW/QFT}.

\subsection{Janus solutions}
\begin{figure}
         \centering
         \begin{subfigure}[b]{0.3\textwidth}
                 \includegraphics[width=\textwidth]{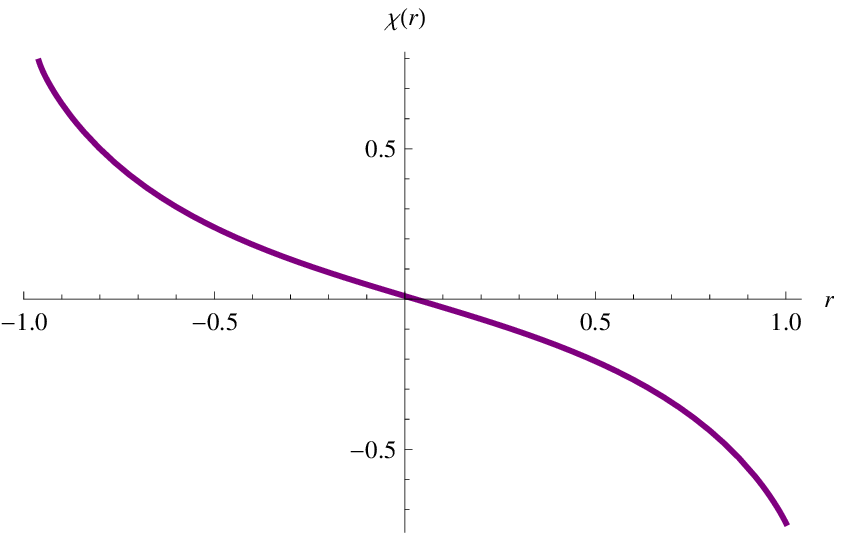}
                 \caption{Solution for $U_1$}
         \end{subfigure}
\begin{subfigure}[b]{0.3\textwidth}
                 \includegraphics[width=\textwidth]{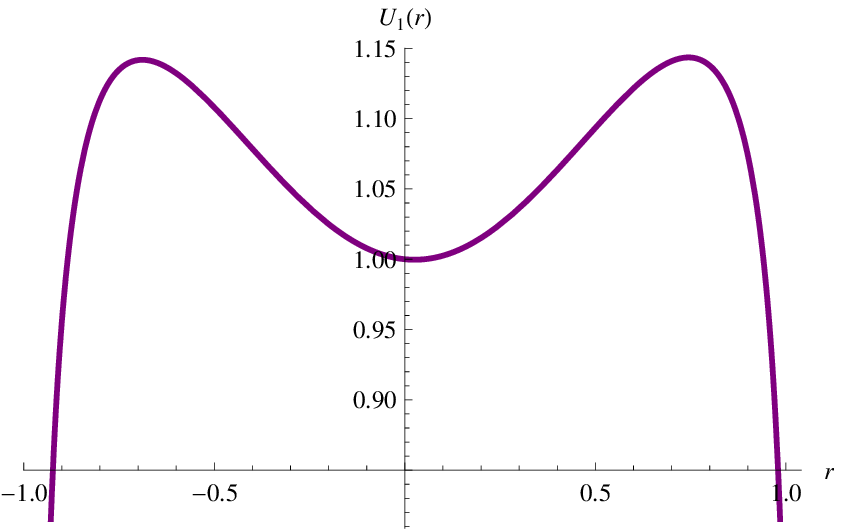}
                 \caption{Solution for $U_1$}
         \end{subfigure}%
         ~ 
         \begin{subfigure}[b]{0.3\textwidth}
                 \includegraphics[width=\textwidth]{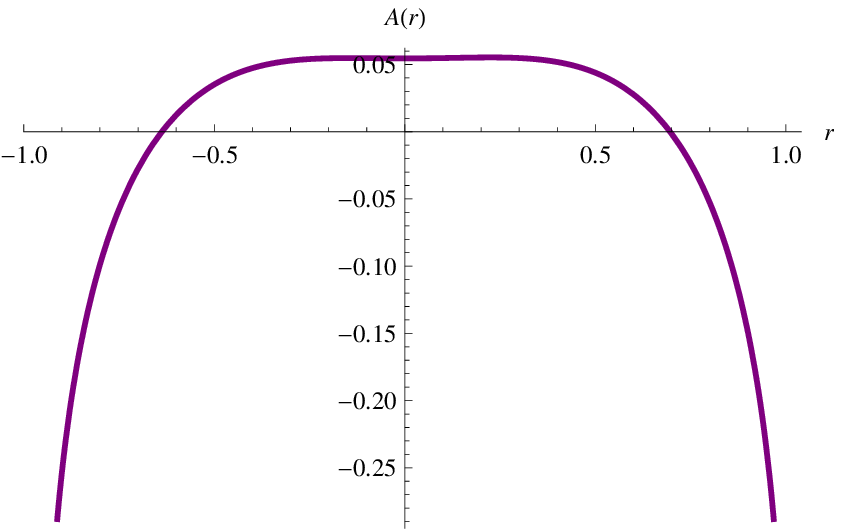}
                 \caption{Solution for $A$}
         \end{subfigure}
         \caption{$N=1$ Janus solution within a truncation to two irrelevant scalars with $k=-1$,
         $\kappa=1$ and $\ell=1$}\label{fig3}
 \end{figure}

\begin{figure}
         \centering
         \begin{subfigure}[b]{0.3\textwidth}
                 \includegraphics[width=\textwidth]{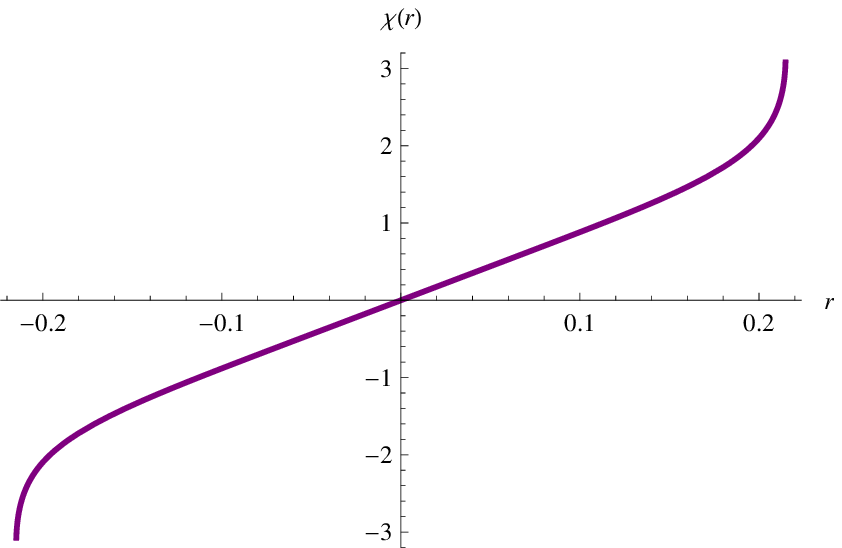}
                 \caption{Solution for $\chi$}
         \end{subfigure}%
         ~ 
         \begin{subfigure}[b]{0.3\textwidth}
                 \includegraphics[width=\textwidth]{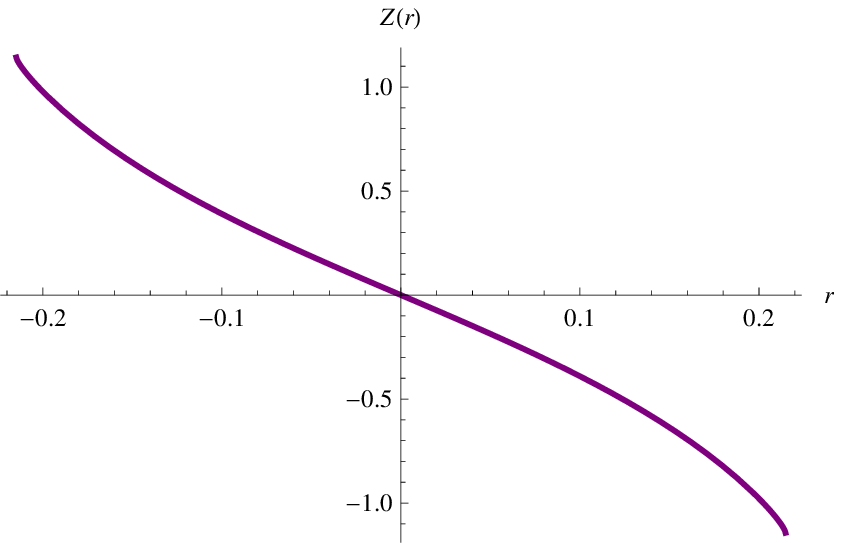}
                 \caption{Solution for $Z$}
         \end{subfigure}
         \begin{subfigure}[b]{0.3\textwidth}
                 \includegraphics[width=\textwidth]{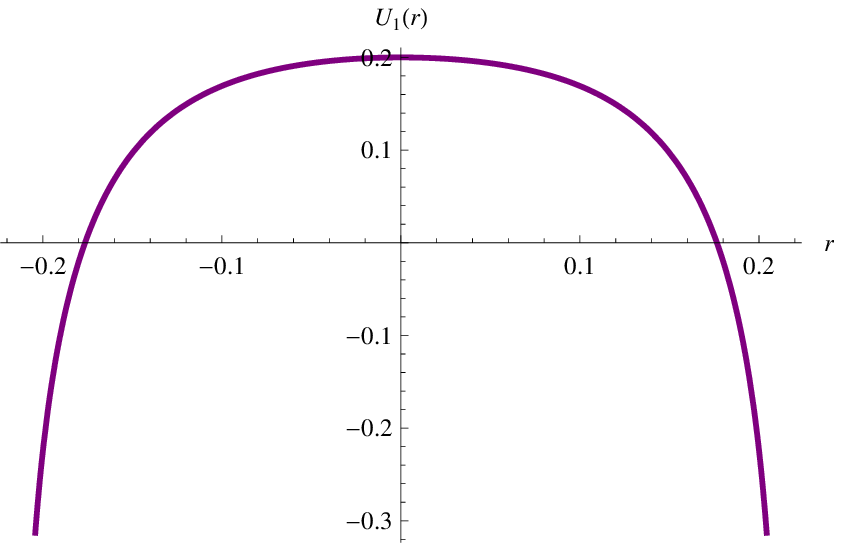}
                 \caption{Solution for $U_1$}
         \end{subfigure}\\
\begin{subfigure}[b]{0.3\textwidth}
                 \includegraphics[width=\textwidth]{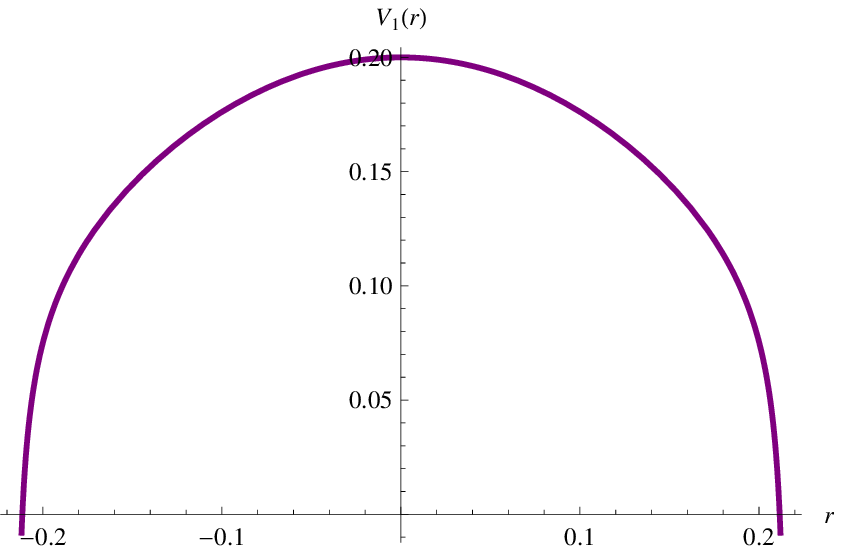}
                 \caption{Solution for $V_1$}
         \end{subfigure}%
         ~ 
         \begin{subfigure}[b]{0.3\textwidth}
                 \includegraphics[width=\textwidth]{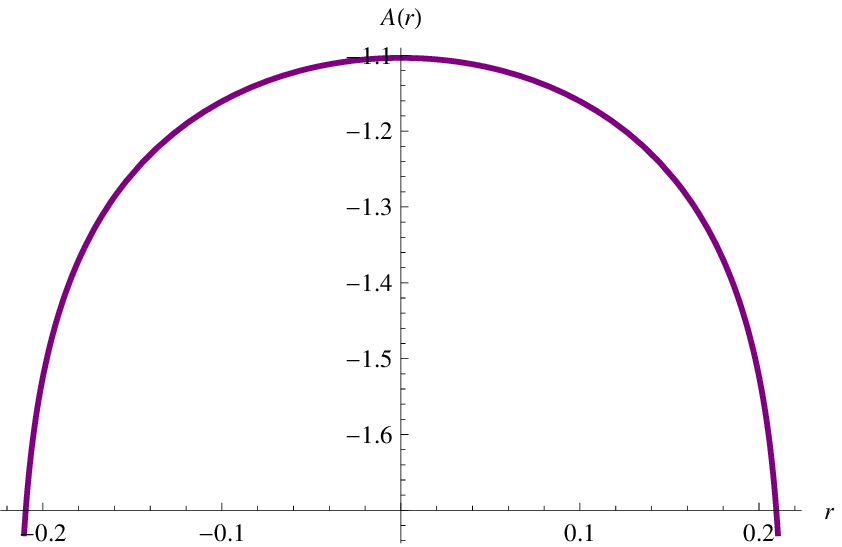}
                 \caption{Solution for $A$}
         \end{subfigure}
         \caption{$N=1$ Janus solution with all $SO(3)$ singlet scalars and $k=-1$, $\kappa=1$ and $\ell=1$}\label{fig4}
 \end{figure}

In the case of $N=1$ supersymmetry, it is possible to have a
supersymmetric Janus solution describing a conformal interface
within the three-dimensional $N=1$ SCFT. The resulting BPS equations
for an $AdS_3$-sliced domain wall metric can be written as
\begin{eqnarray}
U_1'&=&-\frac{2}{3}\frac{A'}{W_1}\frac{\pd W_1}{\pd U_1}-\frac{2}{3}\kappa e^{U_1}\left(\frac{e^{-A}}{\ell W_1}\right)\frac{\pd W_1}{\pd Z},\\
Z'&=&-\frac{2}{3}\frac{A'}{W_1}e^{2U_1}\frac{\pd W_1}{\pd Z}+\frac{2}{3}\kappa e^{U_1}\left(\frac{e^{-A}}{\ell W_1}\right)\frac{\pd W_1}{\pd U_1},\\
V_1'&=&-\frac{4}{9}\frac{A'}{W_1}\frac{\pd W_1}{\pd V_1}-\frac{4}{3}\kappa e^{3V_1}\left(\frac{e^{-A}}{\ell W_1}\right)\frac{\pd W_1}{\pd \chi},\\
\chi'&=&-4e^{6V_1}\frac{A'}{W_1}\frac{\pd W_1}{\pd \chi}+\frac{4}{3}\kappa e^{3V_1}\left(\frac{e^{-A}}{\ell W_1}\right)\frac{\pd W_1}{\pd V_1},\\
{A'}^2&=&W_1^2-\frac{e^{-2A}}{\ell^2}\, .
\end{eqnarray}
These equations reduce to the RG flow equations in the limit
$\ell\rightarrow \infty$ as expected. They take a very similar form
to the equations studied within the $N=8$ and $N=3$ gauged
supergravities in \cite{warner_Janus} and \cite{N3_Janus}. All of
these equations satisfy the corresponding second-order field
equations. We will not present the explicit form of these equations here
due to their complexity. This can be obtained from the above
equations by taking the superpotential $W_1$ from equation
\eqref{N1_superpotential}.
\\
\indent There is however a consistent truncation that can be performed by keeping only the irrelevant deformations. It can be straightforwardly checked that setting $V_1=\frac{U_1}{3}-\frac{1}{3}\ln 5$ and $Z=5\chi$ is a consistent truncation for both the above BPS equations and the corresponding field equations. The resulting equations are given by
\begin{eqnarray}
U_1'&=&\frac{2A'}{\mc{Y}}\left[21e^{4U_1}+50ke^{2U_1}+25k^2+1750\chi^2(e^{2U_1}-k)+30625\chi^4\right]\nonumber \\
& &-\frac{40\kappa \chi e^{-A+U_1}}{\ell \mc{Y}}(7e^{2U_1}+175\chi^2-5k),\\
\chi'&=&-\frac{2\kappa e^{-A+U_1}}{5\ell\mc{Y}}\left[21e^{4U_1}+50e^{2U_1}+25k^2+1750\chi^2(e^{2U_1}-k)+30625\chi^4\right]\nonumber
\\
& &-\frac{8A'\chi e^{2U_1}}{\mc{Y}}(7e^{2U_1}+175\chi^2-5k),\\
A'^2&=&\frac{9}{20}e^{-7U_1}\left[(7e^{U_1}+5k^2)^2+350\chi^2(7e^{2U_1}-5k)+30625\chi^4\right]
-\frac{e^{-2A}}{\ell^2}
\end{eqnarray}
where $\mc{Y}$ is defined by
\begin{equation}
\mc{Y}=(7e^{U_1}+5k^2)^2+350\chi^2(7e^{2U_1}-5k)+30625\chi^4\, .
\end{equation}
Even within this simpler truncation, it is not possible to find any analytic solutions.
\\
\indent We now return to the BPS equations for all $SO(3)$ singlet scalars. As in the $N=8$ gauged supergravity case, these BPS
equations have a turning point at which $A'=0$. Also, the
regular Janus solution is required to approach the $N=1$ $AdS_4$
critical point as $r\rightarrow \pm \infty$. As discussed in
\cite{warner_Janus}, for a given branch of $A'$ near one of these
limits, the first term in the scalar flow equations dominates. When the
solution moves from the critical point, the second term will make
the solution begin to loop around. At the point when $A'=0$, the
other branch of $A'$ equation will bring the solution back to the
$AdS_4$ critical point. The solution preserves $N=(1,0)$ or
$N=(0,1)$ supersymmetry on the two dimensional-interface depending
on the sign of $\kappa$.
\\
\indent However, from an intensive numerical search, we have not
found this type of solutions even starting from $A''>0$ at the
turning point. All of the solutions we obtain are singular on both
sides of the turning point. Example of these solutions for the
two-scalar truncation and all four scalars are shown respectively
in figures \ref{fig3} and \ref{fig4}. Note that the singularities
appearing at both ends correspond to the non-conformal phases of the
dual $N=1$ SCFT studied in the previous section. These are also
physical singularities according to the criterion of
\cite{Maldacena_Nunez_nogo}. Therefore, we expect that these
singular solutions might give some physical insight to the dual
$N=1$ field theories.

\section{Conclusions}\label{conclusions}
We have studied $N=4$ gauged supergravity in four dimensions with
$SO(3)\ltimes (\mathbf{T}^3,\hat{\mathbf{T}}^3)$ gauge group. This
theory is a consistent truncation of eleven-dimensional supergravity
on a tri-sasakian manifold including massive Kaluza-Klein modes. The
theory admits two supersymmetric $AdS_4$ critical points with $N=3$
and $N=1$ supersymmetries and unbroken $SO(3)_R$ R-symmetry. We have
fully analyzed the BPS equations for both cases and checked that
they satisfy all the second-order field equations. This analysis has
not been carried out in the truncation given in
\cite{N010_truncation_Cassani} in which only the structure of the
supermultiplets has been discussed. The result obtained in this
paper is consistent with all the expectations in
\cite{N010_truncation_Cassani} and in a sense could be viewed as an
extension of the analysis in \cite{N010_truncation_Cassani} to
include the fermionic supersymmetry variations.
\\
\indent We have subsequently used these BPS equations to study
possible sueprsymmetric deformations of the dual three-dimensional
$N=3$ and $N=1$ SCFTs. These deformations correspond to turning on
scalar composite operators dual to the massive gravitino multiplet
of the gauged supergravity or their vacuum expectation values. We
have studied a number of RG flows between these SCFTs and
non-conformal field theories in three dimensions. Many of these
deformations lead to various singularities corresponding to possible
non-conformal phases of the dual SCFTs. We have also checked that
all of the new $N=1$ flow solutions presented here flow to physical
singularities. Among the various solutions found in this paper, we
have recovered the $N=3$ flow from $E^{1,2}\times HK$ to
$AdS_4\times N^{010}$ and the $N=1$ flow from $E^{1,2}\times
Spin(7)$ to $AdS_4\times \tilde{S}^7$ studied in
\cite{F4_nunezAdS6}.
\\
\indent The results given here provide additional gravity solutions
to AdS$_4$/CFT$_3$ correspondence and might be useful in many
studies along this line. In addition, we have found a number of
supersymmetric domain wall solutions which might be useful in the
context of DW/QFT correspondence. All of these solutions can be
straightforwardly uplifted to eleven dimensions. The corresponding
prescription of the uplift has also been given. It could be
interesting to further study the implications of these solutions in
the dual $N=1$ SCFT and $N=1$ gauge theory. The interpretation of
these solutions in terms of M-brane geometries when uplifted to
eleven dimensions also deserves further investigation.
\\
\indent Furthermore, we have looked at possible supersymmetric Janus
solutions. In the $N=3$ case, this type of solutions is not possible
at least with unbroken $SO(3)_R$ symmetry. This is similar to the
five-dimensional Janus solution with unbroken $SO(6)$ symmetry
\cite{5D_Janus_classification}. There could also be
non-supersymmetric Janus solutions in this case as well. For the
$N=1$ case, the supersymmetric Janus solution is possible
numerically. This solution corresponds to a two-dimensional
conformal interface with $N=(1,0)$ unbroken supersymmetry. We have
given examples of numerical Janus solutions between $N=1$
non-conformal phases of three-dimensional SCFTs. These solutions
might be useful in the context of interfaced and boundary CFTs
\cite{ICFT_BCFT}. It would be interesting (if possible) to look for
regular Janus solutions interpolating between $N=1$ $AdS_4$ critical
points which describe defected CFTs in three dimensions \cite{dCFT}.
\\
\indent We end the paper by pointing out other possible future
works. First of all, it is interesting to consider more general
solutions with a residual symmetry less than $SO(3)_R$. From the
$N=1$ BPS equations studied here, it could be readily seen that this
analysis would be very complicated. Alternatively, we could consider
solutions with non-vanishing gauge fields that interpolate between
$N=1,3$ $AdS_4$ solutions to $AdS_2\times \Sigma_2$ in which
$\Sigma_2$ is a Riemann surface. These solutions should correspond
to twisted three-dimensional SCFTs and would be interesting in the
study of black hole physics. Another issue, which should be of much
interest, is to construct a more general and complete truncation of
eleven-dimensional supergravity on $N^{010}$. The truncation given
in \cite{N010_truncation_Cassani} has taken into account only
$SU(3)$ singlet fields. This more general truncation could be used
to uplift the RG flows and Janus solutions studied in
\cite{N3_SU2_SU3} and \cite{N3_Janus} resulting in new holographic
solutions in eleven dimensions. Finally, by taking the Betti
multiplet into account, it would be interesting to study baryon
states corresponding to M5-branes wrapped on supersymmetric
$5$-cycles of $N^{010}$ similar to the study of the four-dimensional
gauge theory in \cite{Gubser_4D_baryon}.
\vspace{0.5cm}\\
{\large{\textbf{Acknowledgement}}} \\
The author is very much indebted to Davide Cassani for various
useful correspondences and clarifications on the tri-sasakian
truncation. He would also like to thank Hamburg University for
hospitality while some parts of this work have been done. Many
discussions with Carlos Nunez are gratefully acknowledged. This work
is partially supported by the German Science Foundation (DFG) under
the Collaborative Research Center (SFB) 676 ``Particles, Strings and
the Early Universe'' and The Thailand Research Fund (TRF) under
grant RSA5980037.
\appendix
\section{Field equations for $SO(3)_R$ singlet
scalars}\label{field_eq} In this appendix, we explicitly give the
field equations for all of the four $SO(3)_R$ singlet scalars and
the corresponding Einstein equations. Since the equations in the RG
flow case can be obtained from those of the Janus solutions, we will
only give the equations for the Janus solutions.
\\
\indent The scalar equations are given by
\begin{eqnarray}
U_1''+3A'U_1'-e^{-6U_1-3V_1}\left[2e^{4U_1}+24e^{3U_1+3V_1}-8e^{2U_1+6V_1}-18k^2
\right. & & \nonumber
\\
-4Z^2(2e^{2U_1}+18e^{6V_1}-9k)-18Z^4-8\chi^2(e^{2U_1}+9Z^2)&
&\nonumber
\\
\left.-8\chi Z (2e^{2U_1}-9k+9Z^2)-Z'^2e^{4U_1+3V_1}\right]&=&0,\\
V_1''+3A'V_1'-\frac{1}{3}e^{-6U_1-6V_1}\left[6e^{4U_1+3V_1}+12e^{2U_1+9V_1}-18k^2e^{3V_1}
\right.& &\nonumber \\
+12e^{3V_1}Z^2(6e^{6V_1}-e^{2U_1}+3k)-24\chi Z
e^{3V_1}(e^{2U_1}-3k+3Z^2) & &\nonumber \\
\left.-18Z^4e^{3V_1}-12\chi^2e^{3V_1}(e^{2U_1}+6Z^2)-\chi'^2e^{6U_1}\right]&=&0,\\
\chi''+3A'\chi'-6\chi'V_1'-6e^{-6U_1+3V_1}\left[4e^{2U_1}(\chi+Z)\right.&
&\nonumber \\
\left.+12Z(Z^2+2\chi Z-k)\right]&=&0,\\
Z''+3A'Z'-2U_1'Z'-4e^{-4U_1-3V_1}\left[Z(e^{2U_1}+6e^{6V_1}-3k+3Z^2)\right.&
&\nonumber
\\
\left.+6\chi^2Z+\chi(e^{2U_1}-3k+9Z^2)\right]&=&0\, .
\end{eqnarray}
With the metric ansatz \eqref{DW_ansatz}, the Einstein equations
give rise to the following (dependent) equations
\begin{eqnarray}
2A''+3{A'}^2+\frac{e^{-2A}}{\ell^2}
+\frac{3}{2}{U_1'}^2+\frac{9}{4}{V'_1}^2+\frac{1}{4}e^{-6V_1}{\chi'}^2+\frac{3}{2}e^{-2U_1}{Z'}^2
+V&=&0,\,\,\,\,\\
3{A'}^2+\frac{3}{\ell^2}e^{-2A}-\frac{3}{2}{U_1'}^2
-\frac{9}{4}{V_1'}^2-\frac{1}{4}e^{-6V_1}{\chi'}^2-\frac{3}{2}e^{-2U_1}{Z'}^2
+V&=&0
\end{eqnarray}
where $V$ is the scalar potential given in \eqref{Scalar_potential}.

\section{BPS equations for $N=1$ supersymmetry}\label{N1_BPS_eq}
We give the BPS equations for the $N=1$
RG flow solutions here. These equations are given by
\begin{eqnarray}
U_1'&=&\frac{e^{-\frac{3}{2}(2U_1+V_1)}}{\mc{Q}}\left[(e^{2U_1}+2e^{U_1+3V_1}+k)(e^{2U_1}+4e^{U_1+3V_1}+3k)\right.\nonumber
\\
&
&+2Z^2(2e^{2U_1}+6e^{6V_1}+5e^{U_1+3V_1}-3k)+4\chi^2(2e^{2U_1}+3Z^2)\nonumber
\\
& &\left.+3Z^4+4\chi Z (2e^{2U_1}+3Z^2-3k)\right],\\
V_1'&=&\frac{e^{-\frac{3}{2}(2U_1+V_1)}}{\mc{Q}}\left[(e^{2U_1}+k)^2-4e^{2U_1+6V_1}+2Z^2(e^{2U_1}-2e^{6V_1}-k)\right.
\nonumber
\\
& &\left.+Z^4+4\chi^2(e^{2U_1}+Z^2)+4\chi Z
(e^{2U_1}+Z^2-k)\right],\\
Z'&=&-\frac{2e^{-U_1-\frac{3}{2}V_1}}{\mc{Q}}\left[2Ze^{6V_1}+2Ze^{U_1+3V_1}+(\chi+Z)(e^{2U_1}+2\chi
Z+Z^2-k)\right],\nonumber \\
& &\\
\chi'&=&-\frac{12e^{-3U_1+\frac{9}{2}V_1}}{\mc{Q}}\left[(2\chi+Z)e^{2U_1}+Z(Z^2+2\chi
Z-k)\right]
\end{eqnarray}
where
\begin{equation}
\mc{Q}=\sqrt{\left[2e^{3V_1}Z+2(\chi+Z)e^{U_1}\right]^2+\left[e^{2U_1}+2e^{U_1+3V_1}+k-Z(2\chi+Z)\right]^2}\,
.
\end{equation}


\end{document}